\documentclass{article}

\usepackage{graphics}
\usepackage{graphicx}
\usepackage{latexsym}
\usepackage{amsfonts}
\usepackage{amssymb}

\usepackage{amsmath}       
\usepackage{url}
\usepackage{subfigure}

\begin{document}

\title{Logical Gates via Gliders Collisions}

\author{Genaro J. Mart\'{\i}nez$^{1,2}$, Andrew Adamatzky$^{2}$ \\ Kenichi Morita$^{3}$}

\date{14 March 2018\footnote{Published in: {\em Reversibility and Universality: Essays Presented to Kenichi Morita on the Occasion of his 70th Birthday}, A. Adamatzky (ed.), chapter 9, pages 199--220. Springer, 2018. A short version is published as: ``Conservative Computing in a One-dimensional Cellular Automaton with Memory,'' {\em Journal of Cellular Automata} {\bf 13(4)} 325--346, 2018.}}


\maketitle

\begin{centering}
$^1$ Escuela Superior de C\'omputo, Instituto Polit\'ecnico Nacional, M\'exico \\
$^2$ Unconventional Computing Lab, University of the West of England, Bristol, United Kingdom \\
$^3$ Hiroshima University, Higashi Hiroshima, Japan \\
\end{centering}

\begin{abstract}
\noindent
 An elementary cellular automaton with memory is a chain of finite state machines (cells) updating their state simultaneously and by the same rule. Each cell updates its current state depending on current states of its immediate neighbours and a certain number of its own past states. Some cell-state transition rules support gliders, compact patterns of non-quiescent states translating along the chain. We present designs of logical gates, including reversible Fredkin gate and controlled {\sc not} gate, implemented via collisions between gliders. 

\end{abstract}


\section{Preliminaries}

When designing a universal cellular automata (CA) we aim, similarly to designing small universal Turing machines, to minimise the number of cell states, size of neighbourhood and sizes of global configurations involved in a computation \cite{kn:CS11, kn:Hey98, kn:Ada01, kn:ACA05, kn:Ada10, kn:Hutt10, kn:Ren16, kn:Mit01, kn:Mills08, kn:von66}. History of 1D universal CA is long yet exciting.\footnote{Complex Cellular Automata Repository \url{http://uncomp.uwe.ac.uk/genaro/Complex_CA_repository.html}} In 1971 Smith III proved that a CA for which a number of cell-states multiplied by a neighbourhood size equals 36 simulates a Turing machine \cite{kn:Smi71}. Sixteen years later Albert and Culik II designed a universal 1D CA with just 14 states and  totalistic cell-state transition function~\cite{kn:AC87}. In 1990 Lindgren and Nordahl  reduced the number of cell-states to 7~\cite{kn:LN90}. These proofs were obtained using signals interaction in CA. Another 1D universal CA employing signals was designed by Kenichi Morita in 2007 \cite{kn:Mor07}: the triangular reversible CA which evolves in partitioned spaces (PCA).

In 1998, Cook demonstrated that elementary CA, i.e. with two cell-states and three-cell neighbourhood, governed by rule 110 is universal~\cite{kn:Cook04, kn:Wolf02}. He did this by  simulating a cyclic tag system in a CA.\footnote{A reproduction of this machine working in rule 110 can be found in \url{http://uncomp.uwe.ac.uk/genaro/rule110/ctsRule110.html}} Operations were implemented with 11 gliders and a glider gun.

We will show that an elementary CA with memory, where every cells updates its state depending not only on its two immediate neighbours but also on its own past states, exhibits gliders which collision dynamics allows for implementation of logical gates just with one glider.  We demonstrate implementation of {\sc not} and {\sc and} gates, {\sc delay}, {\sc nand} gate, {\sc majority} gate, and {\sc cnot} and Fredkin gates.

\section{Elementary Cellular Automata}

One-dimensional elementary CA (ECA)~\cite{kn:Wolf94} can be represented as a 4-tuple $\langle \Sigma,\varphi,\mu,c_0 \rangle$, where $\Sigma=\{0, 1 \}$ is a binary alphabet (cell states), $\varphi$ is a local transitions function, $\mu$ is a cell neighbourhood, $c_0$ is a start configuration. The system evolves on an array of {\it cells} $x_i$, where $i \in Z$ (integer set) and each cell takes a state from the $\Sigma=\{0, 1 \}$. Each cell $x_i$ has three neighbours including itself: $\mu(x_i)=(x_{i-1}, x_i, x_{i+1})$. The array of cells \{$x_i$\} represents a {\it global configuration} $c$, such that $c \in \Sigma^*$. The set of finite configurations of length $n$ is represented as $\Sigma^n$. Cell states in a configuration $c(t)$ are updated to next configuration $c(t+1)$ simultaneously by a the local transition function $\varphi$: $x_i^{t+1}=\varphi(\mu(x_i))$. Evolution of ECA is represented by a sequence of finite configurations $\{c_i\}$ given by the global mapping, $\Phi:\Sigma^n \rightarrow \Sigma^n$.

\section{Elementary Cellular Automaton Rule 22}

Rule 22 is an ECA with the following local function:

\begin{equation}
\varphi_{R22} = \left\{
	\begin{array}{lcl}
		1 & \mbox{if} & 100, 010, 001 \\
		0 & \mbox{if} & 111, 110, 101, 011, 000
	\end{array} \right..
\end{equation}

\begin{figure}
\begin{center}
\subfigure[]{\scalebox{0.57}{\includegraphics{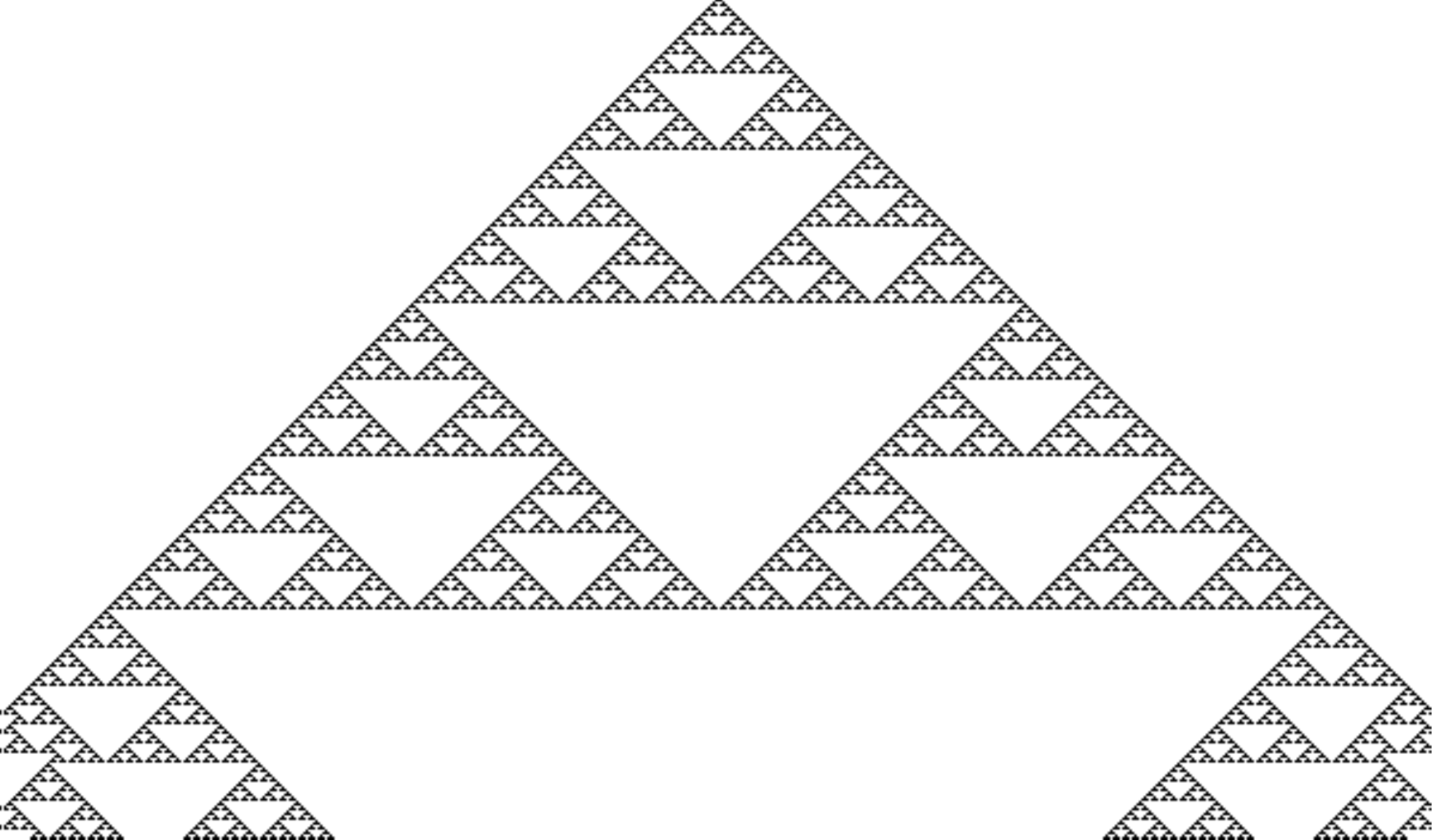}}}
\subfigure[]{\scalebox{0.57}{\includegraphics{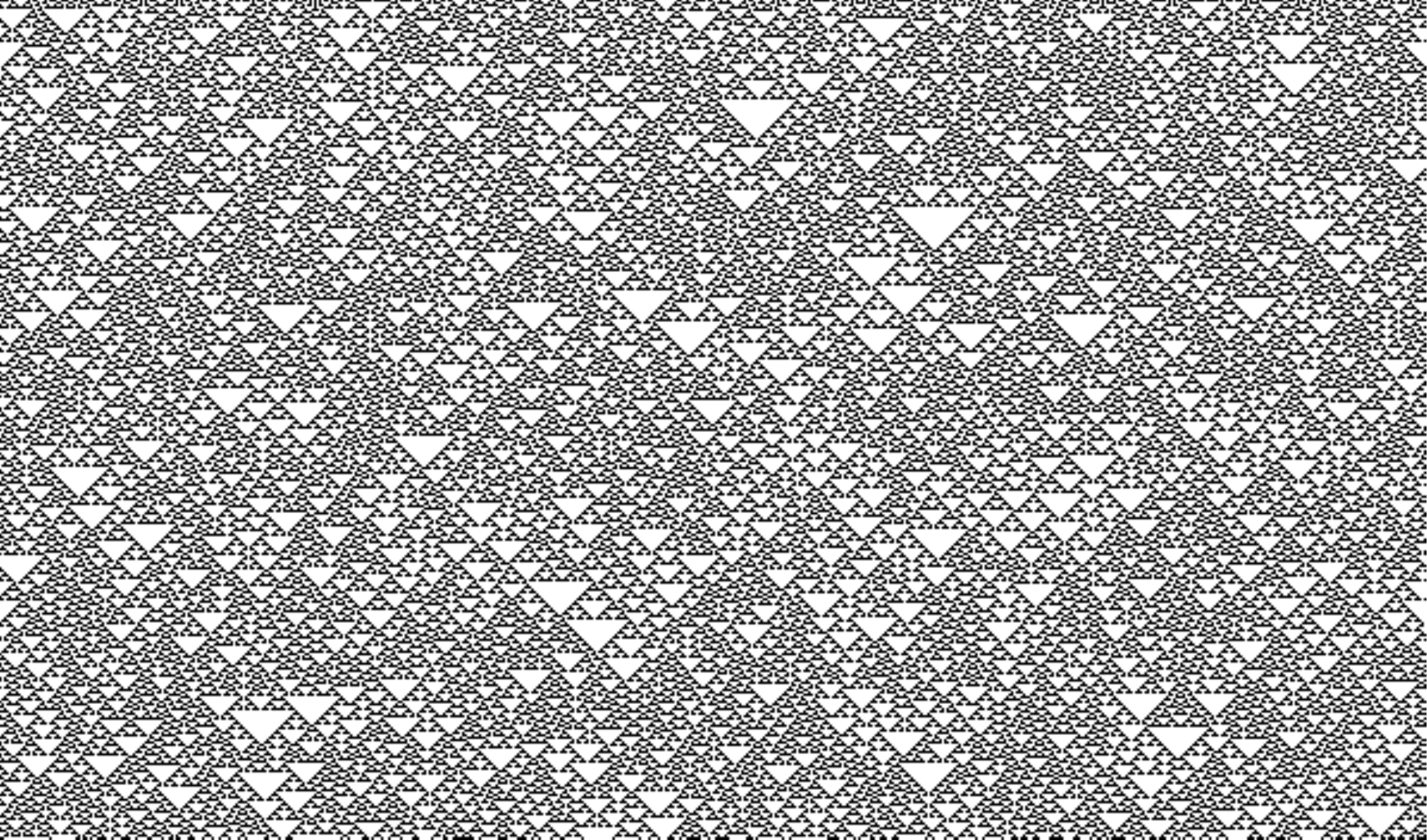}}}
\end{center}
\caption{Typical evolution of ECA rule 22 (a) from a single cell in state `1' and (b) from a random initial configuration where half of the cells, chosen at random, are assigned state `1'. The ECA consists of 598 cells and evolves for 352 generations. White colour represents state `0' and dark colour represents state `1'.}
\label{evolutions22}
\end{figure}

The local function $\varphi_{R22}$ has a probability of 37.5\% to get states 1 in the next generation and much higher probability to get state 0 in the next generation. Examples of evolution of ECA Rule 22 from a single cell in state `1' and from a random configurations are shown in Fig.~\ref{evolutions22}.

\section{Elementary Cellular Automata with Memory}

Conventional CA are ahistoric (memoryless): the new state of a cell depends on the neighbourhood configuration solely at the preceding time step of $\varphi$. CA with {\it memory} (CAM) can be considered as an extension of the standard framework of CA where every cell $x_i$ is allowed to remember some period of its previous evolution. CAM was introduced by Sanz, see overview in~\cite{kn:Alo09}. To implement a memory function we need to specify the kind of memory $\phi$, as follows:

\begin{equation}
\phi (x^{t-\tau}_{i}, \ldots, x^{t-1}_{i}, x^{t}_{i}) \rightarrow s_{i}.
\end{equation}

The parameter $\tau < t$ determines the depth, or a degree, of the memory and each cell $s_{i} \in \Sigma$ being a state function of the series of states of the cell $x_i$ with memory up to time-step. To execute the evolution we apply the original rule as:

\begin{equation}
\varphi(\ldots, s^{t}_{i-1}, s^{t}_{i}, s^{t}_{i+1}, \ldots) \rightarrow x^{t+1}_i.
\end{equation}

The main feature in CAM is that the mapping $\varphi$ remains unaltered, while historic memory of all past iterations is retained by featuring each cell in the context of history of its past states in $\phi$. This way, cells {\it canalise} memory to the map $\varphi$. For example, we can consider memory function $\phi$ as a {\it majority memory} $\phi_{maj} \rightarrow s_{i}$, where in case of a tie given by $\Sigma_1 = \Sigma_0$ in $\phi$ we will take the last value $x_i$. In this case, function $\phi_{maj}$ represents the classic majority function for three values \cite{kn:Mins67} as follows:

\begin{equation}
(a \wedge b) \vee (b \wedge c) \vee (c \wedge a)
\end{equation}

\noindent that represents the cells $(x^{t-\tau}_{i}, \ldots, x^{t-1}_{i}, x^{t}_{i})$ and define a temporal ring of $s$ cells, before reaching the next global configuration $c$.

\section{Elementary Cellular Automaton with Memory Rule $\phi_{R22maj:4}$}

ECA with memory (ECAM) rule $\phi_{R22maj:4}$ employ the majority memory ($maj$) and degree of memory $\tau=4$.

\begin{figure}
\centerline{\includegraphics[width=4.8in]{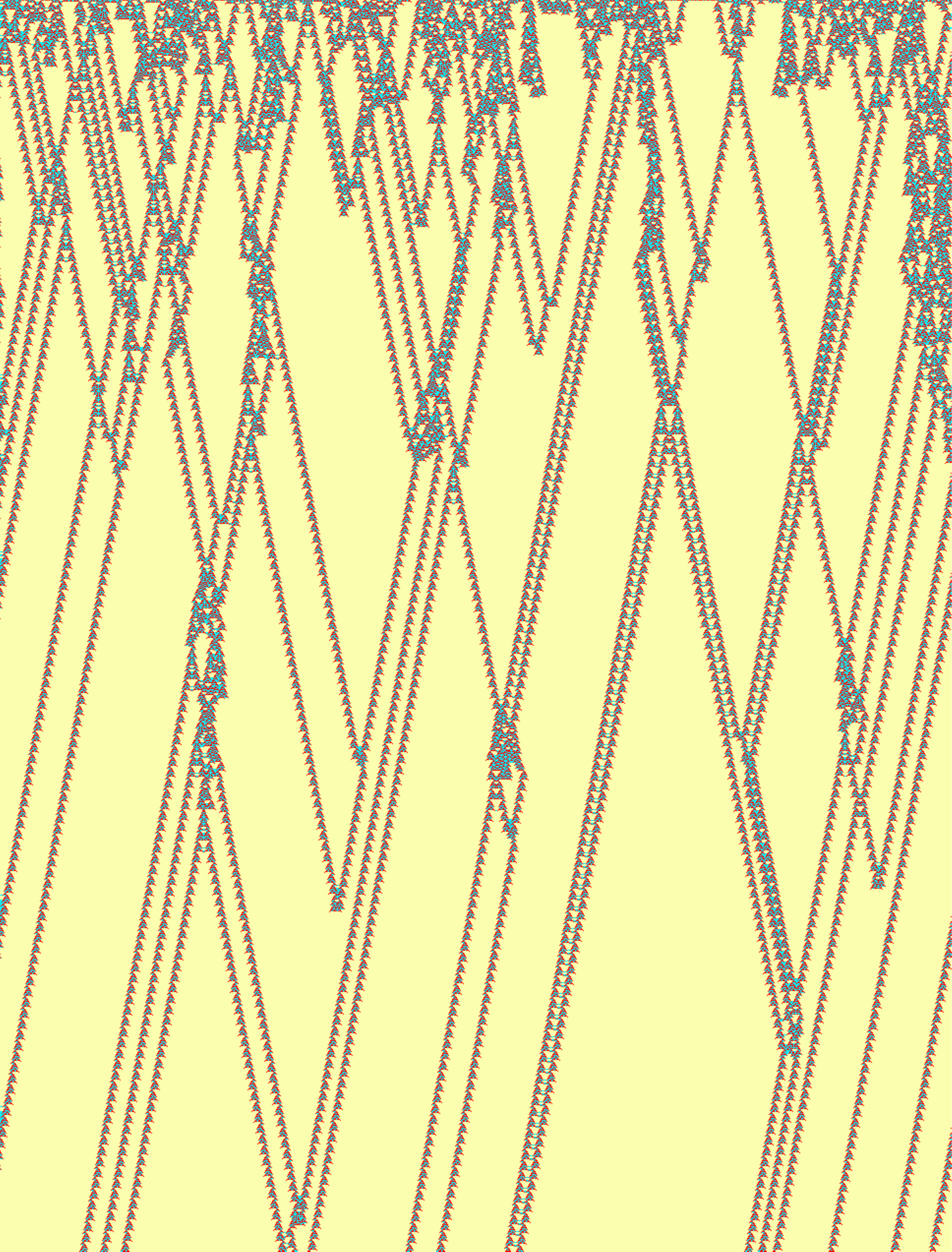}}
\caption{Typical evolution of $\phi_{R22maj:4}$ from a random initial configuration with the ratio of 37\% on a ring of 968 cells for 1,274 generations. Filter and a selection of colors are used to improve the view of particles.}
\label{evolR22maj4}
\end{figure}

Figure~\ref{evolR22maj4} shows a typical evolution of ECAM rule $\phi_{R22maj:4}$ from a random initial condition. There we observe emergence of gliders travelling and colliding with each other.

\begin{table}[th]
\begin{center}
\begin{tabular}{ c | c c c c}
localisation & period & shift & velocity & mass \\
\hline
$g_L$ & 11 & $-2$ & $-2/11$ & 38 \\
$g_R$ & 11 & 2 & $2/11$ & 38 \\
\end{tabular}
\caption{Properties of gliders in ECAM rule $\phi_{R22maj:4}$.}
\label{pgliders}
\end{center}
\end{table}

The set of gliders ${\cal G}_{\phi_{R22maj:4}} = \{g_L,g_R\}$ are defined in a square of $11 \times 11$, their properties are shown in Tab.~\ref{pgliders}.

\begin{figure}[th]
\centerline{\includegraphics[width=4.8in]{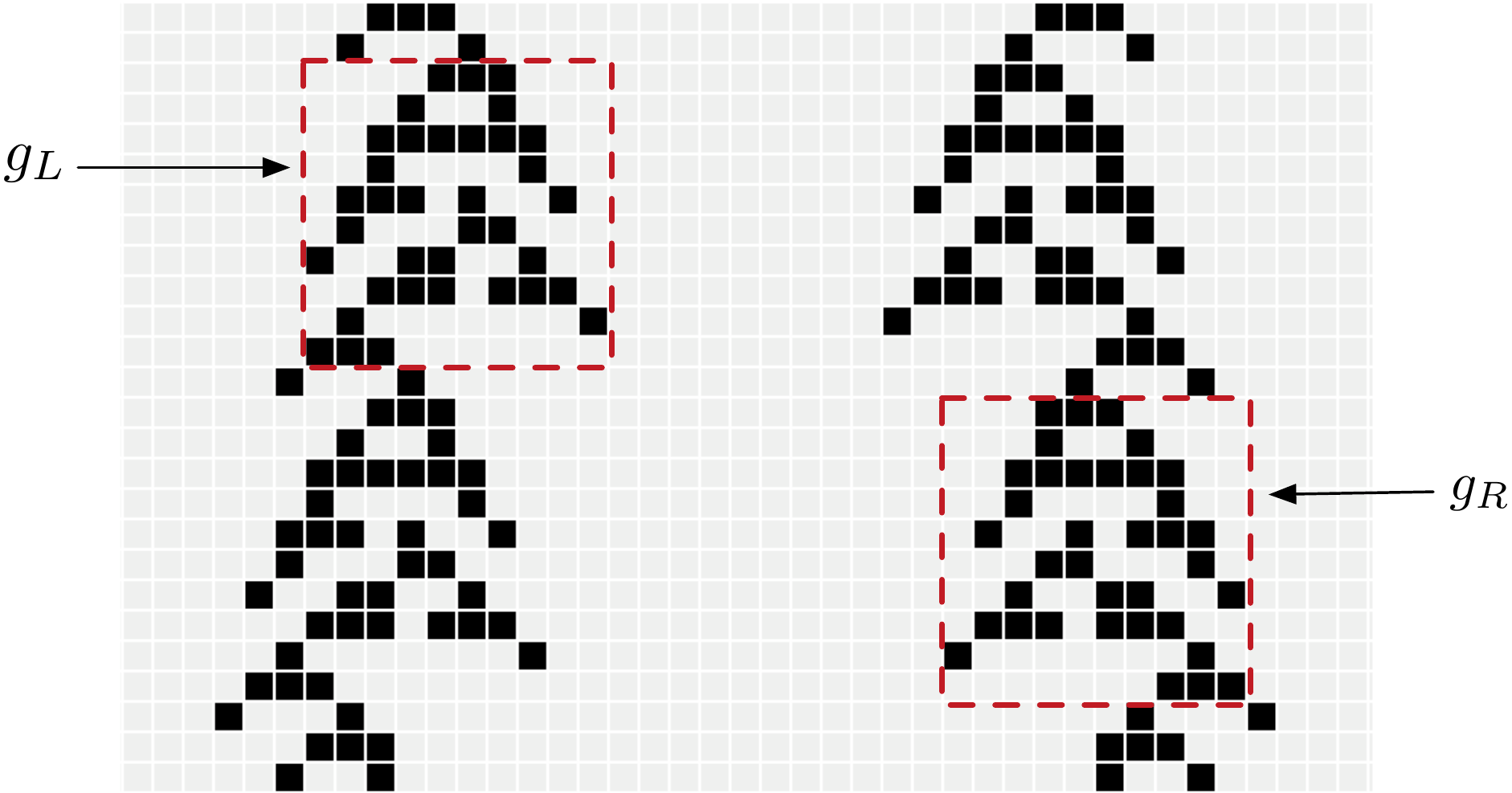}}
\caption{$g_L$ and $g_R$ gliders in ECAM rule $\phi_{R22maj:4}$.}
\label{gliders}
\end{figure}

We undertook a systematic analysis of binary collisions between gliders and found 44 different types of the collision outcomes, we selected some types of the collisions to realise logical gates presented in this paper (see Appendix A).

\section{Logic gates via gliders collisions}

\begin{figure}[th]
\begin{center}
\subfigure[]{\scalebox{0.39}{\includegraphics{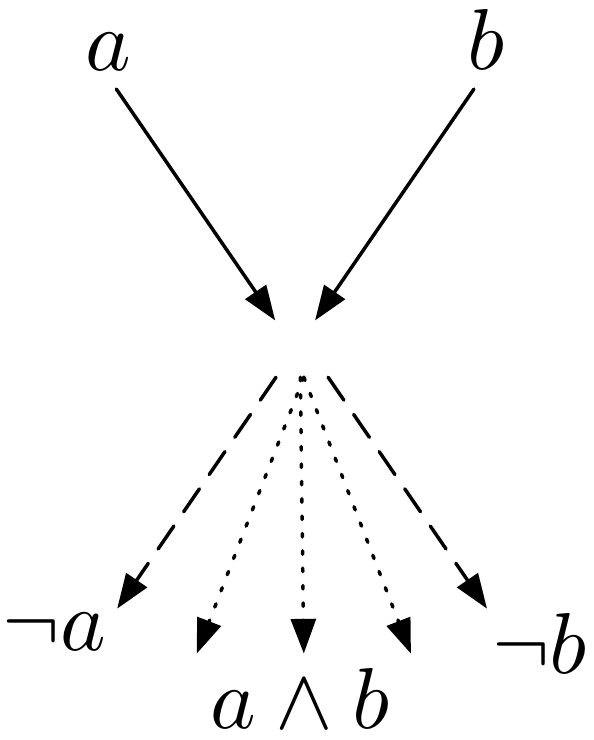}}} 
\subfigure[]{\scalebox{0.39}{\includegraphics{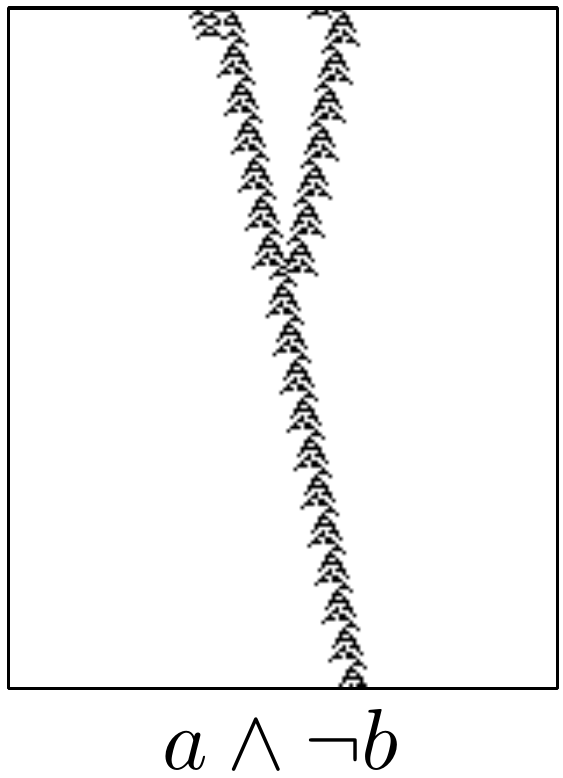}}} 
\subfigure[]{\scalebox{0.39}{\includegraphics{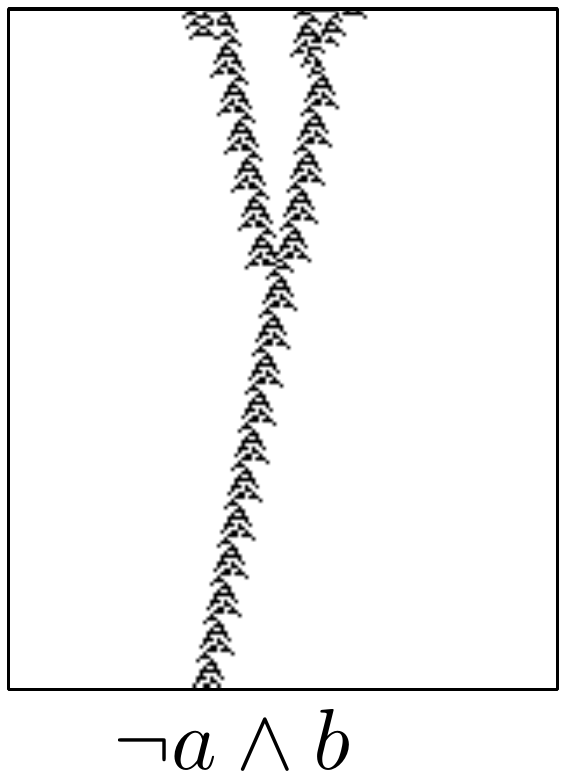}}} 
\subfigure[]{\scalebox{0.39}{\includegraphics{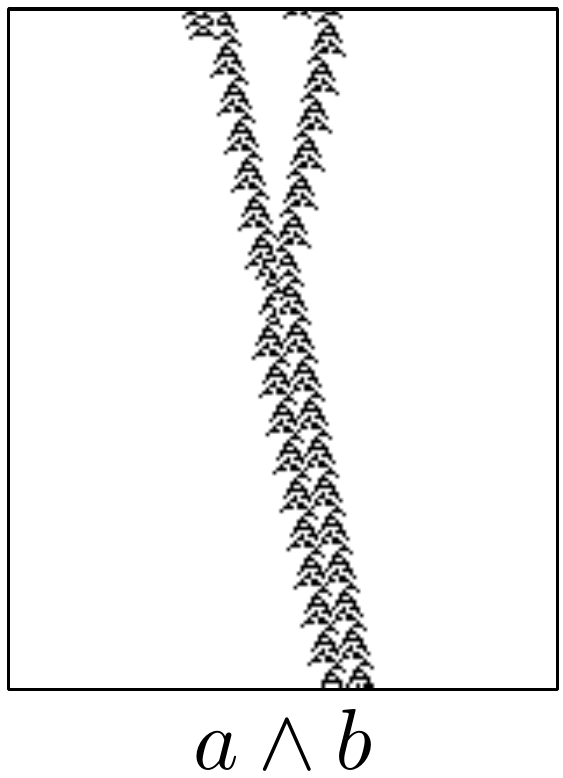}}} 
\subfigure[]{\scalebox{0.39}{\includegraphics{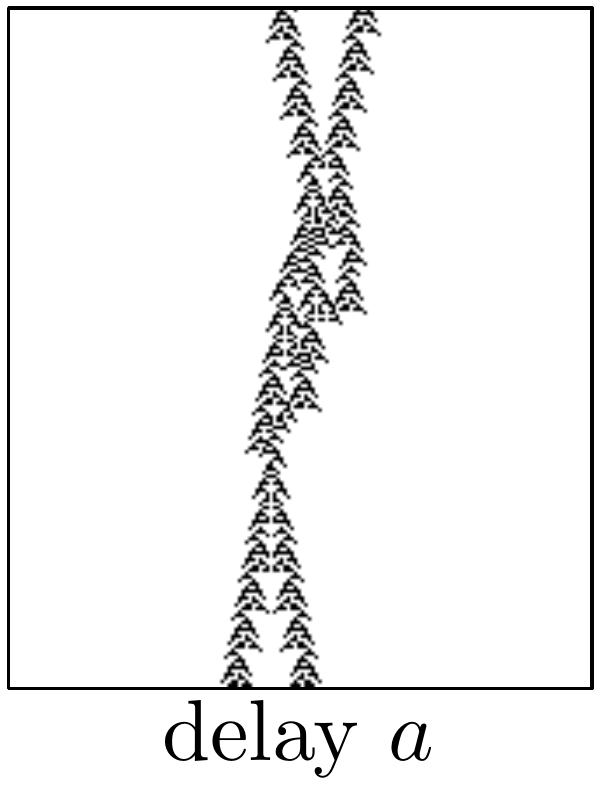}}}
\end{center}
\caption{Examples of logical gates implemented via glider collisions. Presence of a glider is logical {\sc Truth} `1', absence logical {\sc False} `0'. (a) Scheme of the gate. (bc) {\sc and-not} gate, 
(d) {\sc and}, (e)~{\sc delay}}.
\label{basicGates}
\end{figure}

We assume presence of a glider at a given site of space and time is a logical `1' ({\sc True}) and absence is `0' ({\sc False}). Logic gates $f(a,b) \rightarrow \neg a$ $|$ $\neg b$ $|$ $a \wedge b$ are realised via binary collisions of gliders in rule $\phi_{R22maj:4}$. Figure~\ref{basicGates}bc shows how {\sc and-not} gate is realised.  Fig.~\ref{basicGates}d demonstrates implementations of {\sc and} gate. Also a {\sc delay} operator can be implemented by delaying glider  $a$ and conserving momentum of  glider $b$ as shown in Fig.~\ref{basicGates}e 

\begin{figure}
\begin{center}
\subfigure[]{\scalebox{0.9}{\includegraphics{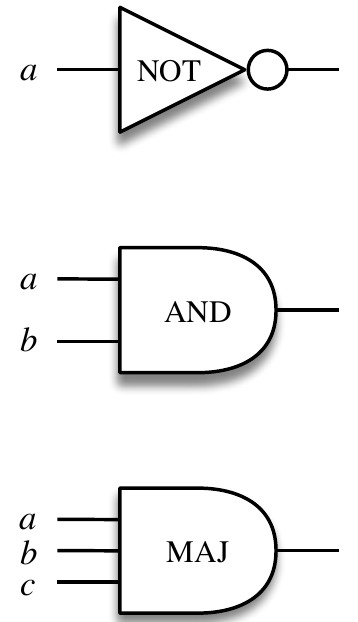}}} \hspace{0.8cm}
\subfigure[]{\scalebox{0.9}{\includegraphics{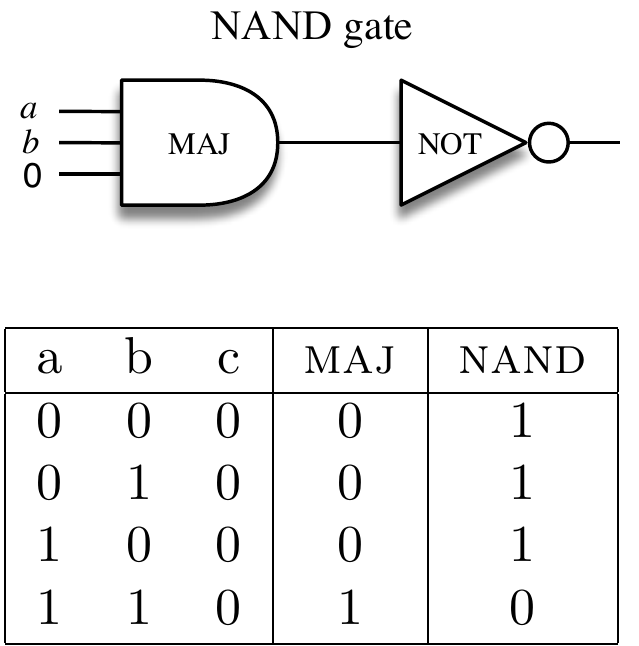}}}
\end{center}
\caption{Gate {\sc nand} made of {\sc majority} and {\sc not} gates. 
(a) Schematics of gates. 
(b) Design of {\sc nand} gate.}
\label{majorityGate-0}
\end{figure}

A {\sc nand} gate can be cascaded from {\sc majority} and {\sc not} gates. To design {\sc nand} gate, we fix third value in {\sc majority} gate to produce {\sc and} gate as illustrated in Fig~\ref{majorityGate-0}b. Later a {\sc not} gate is cascaded to get a {\sc nand} gate (Fig.~\ref{majorityGate-0}a).

\begin{figure}
\begin{center}
\subfigure[]{\scalebox{0.9}{\includegraphics{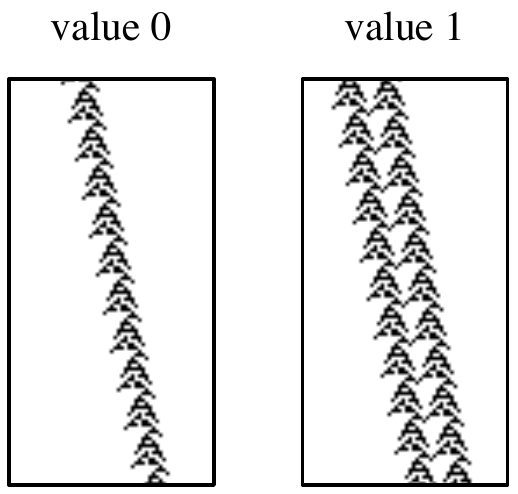}}} \hspace{1.0cm}
\subfigure[]{\scalebox{0.95}{\includegraphics{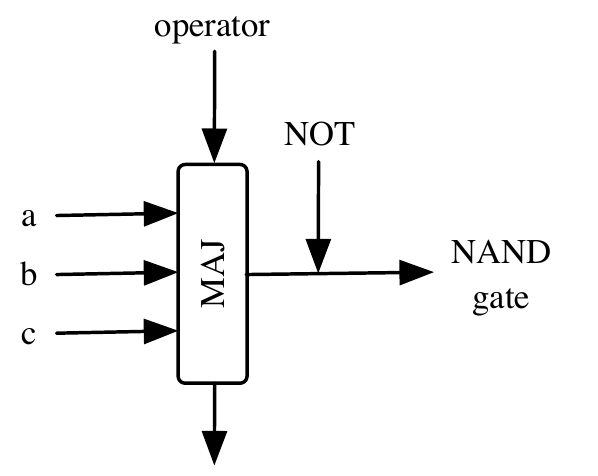}}} \hspace{0.5cm}
\end{center}
\caption{Gate {\sc majority}. (a) Binary values represented by gliders, 
(b) scheme of an {\sc nand} gate using {\sc majority} and {\sc not} gates.}
\label{majorityGate-1}
\end{figure}

Figure~\ref{majorityGate-1}a represents a value `0' as one glider and a value `1' as a couple of gliders. Figure~\ref{majorityGate-1}b presents a scheme of {\sc nand} gate in $\phi_{R22maj:4}$. We have three-input values/gliders that will be evaluated by one control glider travelling perpendicularly to input gliders. The control glider transforms $f(a,b,c)$ into a majority function. Further, other glider acts as an active signal in {\sc not} gate.

\newpage

\begin{figure}
\begin{center}
\subfigure[]{\scalebox{0.85}{\includegraphics{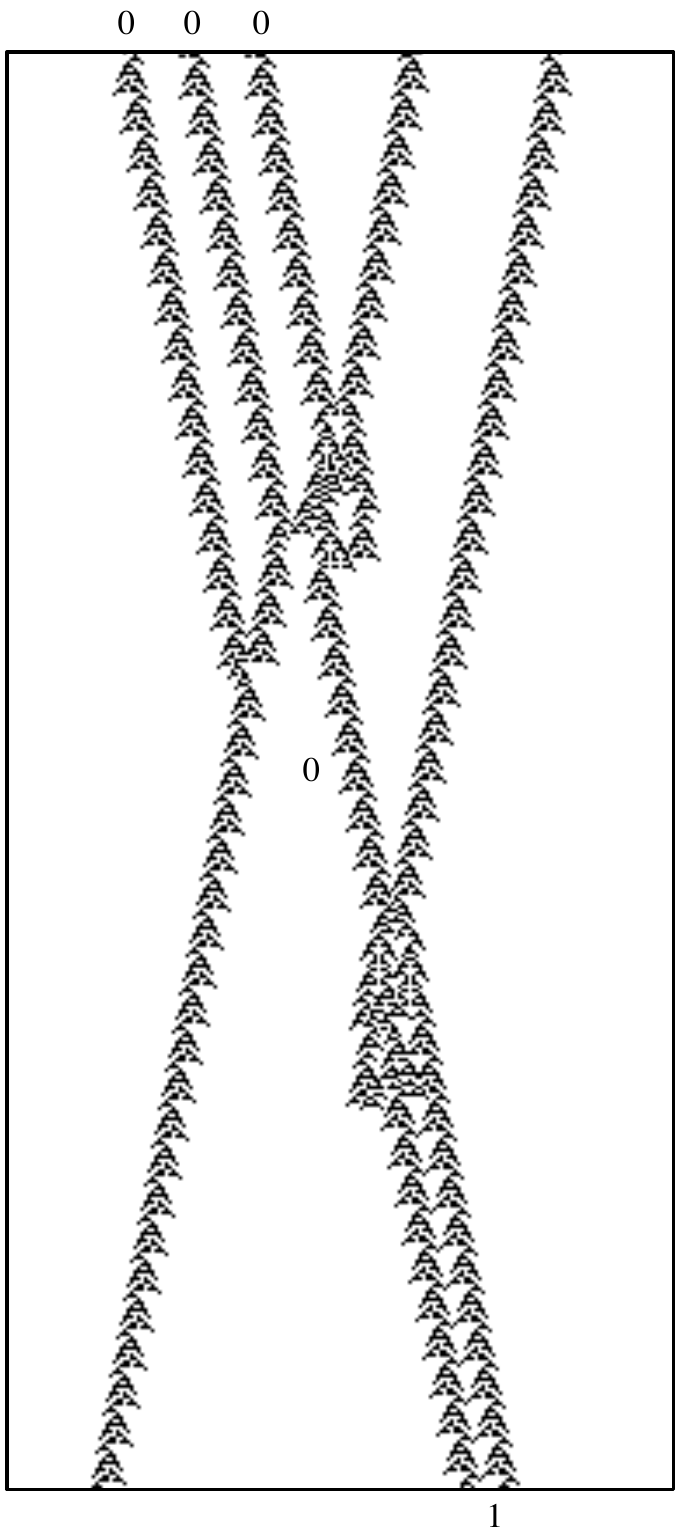}}} \hspace{0.1cm}
\subfigure[]{\scalebox{0.85}{\includegraphics{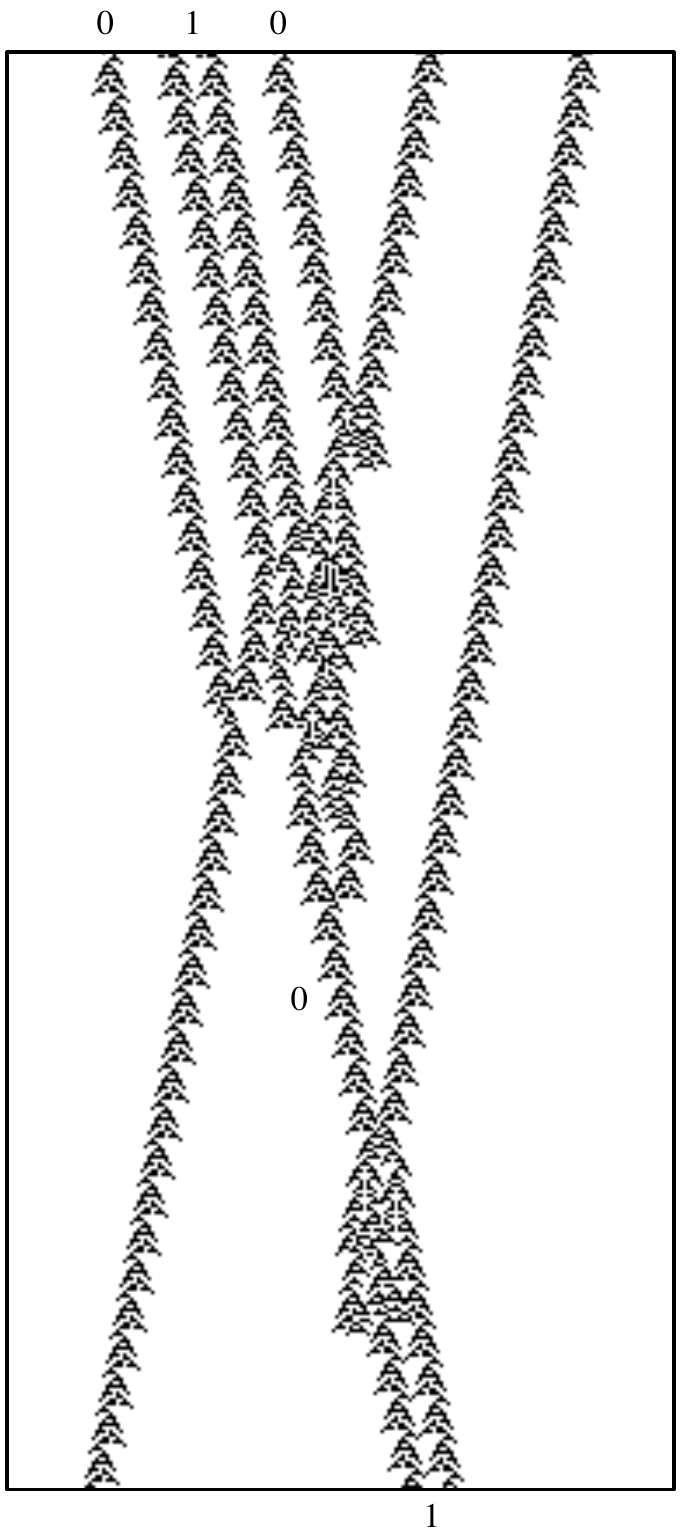}}}
\end{center}
\caption{Gate {\sc nand} implemented with {\sc majority} and {\sc not} gates in ECAM rule $\phi_{R22maj:4}$ following the scheme proposed in Fig.~\ref{majorityGate-1}b. (a) Input values $f(0,0,0)$ and (b) input values $f(0,1,0)$. Intervals between gliders in the initial condition are the same in all operations.}
\label{majorityGate-2}
\end{figure}

\begin{figure}
\begin{center}
\subfigure[]{\scalebox{0.85}{\includegraphics{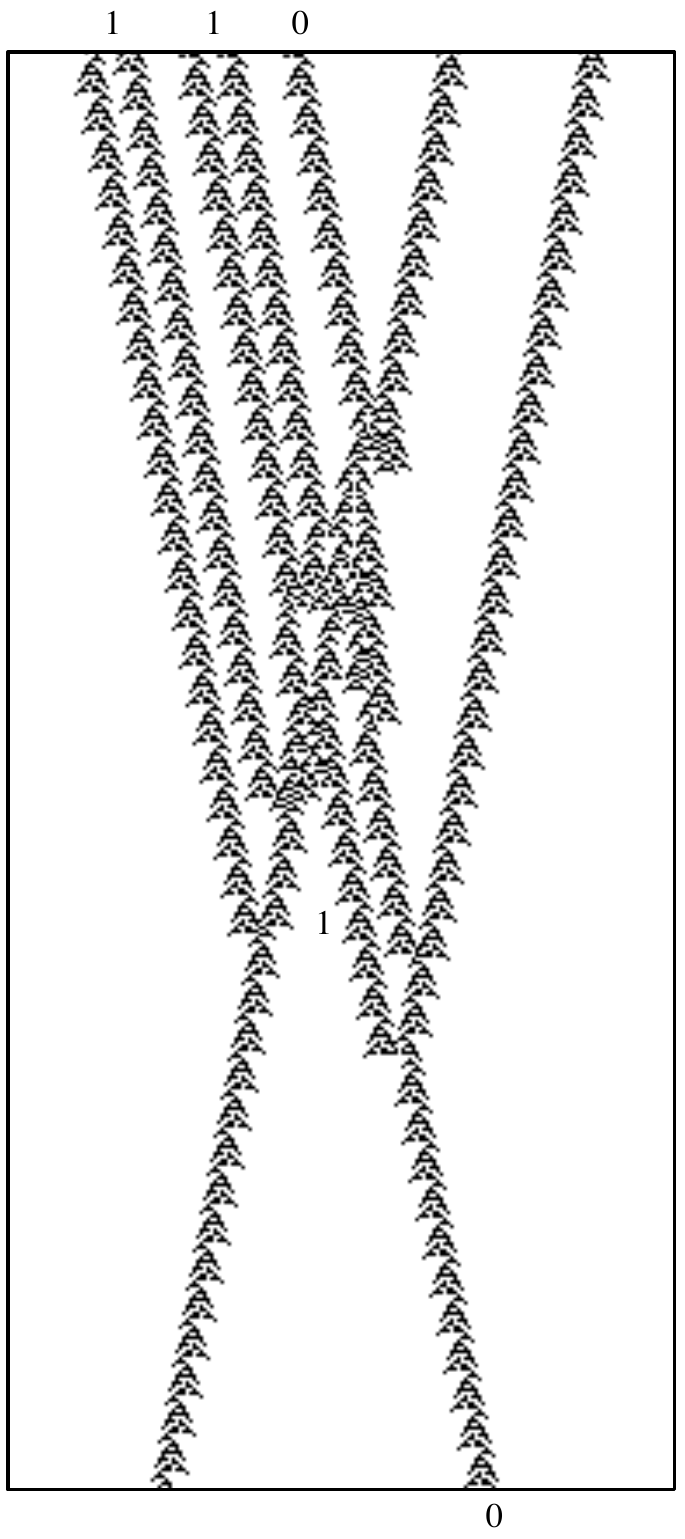}}} \hspace{0.1cm}
\subfigure[]{\scalebox{0.85}{\includegraphics{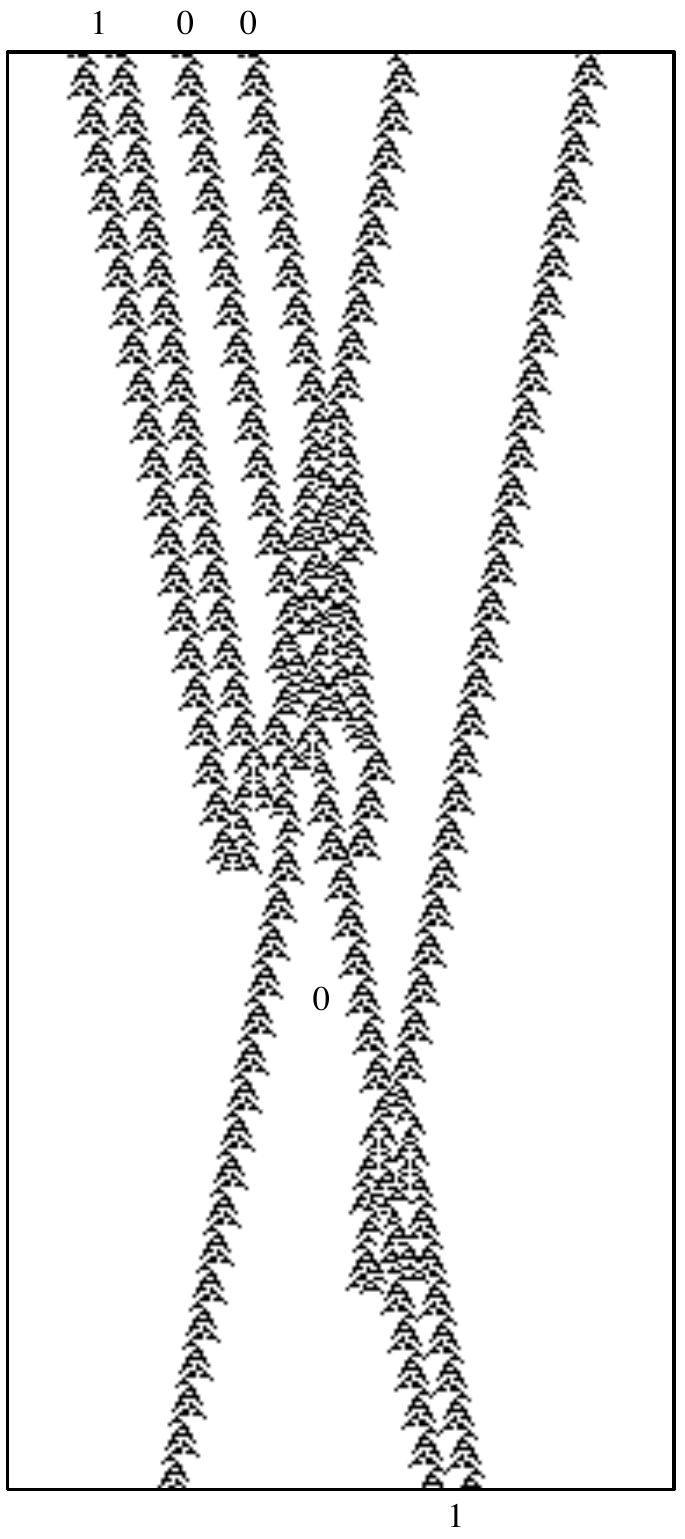}}}
\end{center}
\caption{Gate {\sc nand} implemented with {\sc majority} and {\sc not} gates in ECAM rule $\phi_{R22maj:4}$ following the scheme proposed in Fig.~\ref{majorityGate-1}b. (a) Input values $f(1,1,0)$ and (b) input values $f(1,0,0)$.}
\label{majorityGate-2A}
\end{figure}

Figures~\ref{majorityGate-2} and ~\ref{majorityGate-2A} show the implementation of {\sc nand} gate in $\phi_{R22maj:4}$, encoded in its initial condition set of gliders in six positions. As was shown in the scheme (Fig.~\ref{majorityGate-1}c), gliders coming from the left side represent binary values and a glider coming from the right  acts as a control glider for the values of the {\sc majority} gate. The control glider continues its travel undisturbed, after interaction with input gliders, and therefore it can be recycled in further {\sc majority} gate.

\subsection{Ballistic collisions}

A concept of ballistic logical gates represent Boolean values by {\it balls} which preserve their identity during collision but change their velocity vectors; the balls are routed in the computing space using mirrors~\cite{kn:Toff02}. The {\it billiard ball model} was advanced by Margolus in his designs of partitioned CA to demonstrate a logical universality of a billiard-ball model and to implement Fredkin gate with soft spheres in billiard-ball model~\cite{kn:Marg84, kn:Marg98}.

\begin{figure}[th]
\begin{center}
\subfigure[]{\scalebox{0.63}{\includegraphics{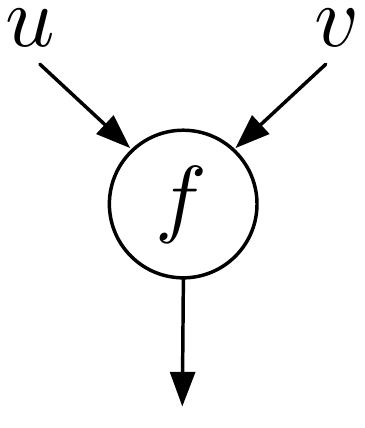}}} \hspace{0.5cm}
\subfigure[]{\scalebox{0.6}{\includegraphics{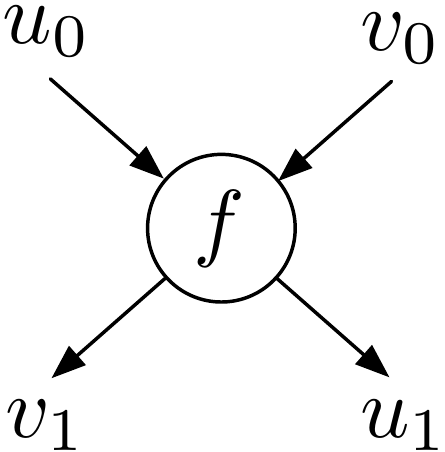}}} \hspace{0.5cm}
\subfigure[]{\scalebox{0.6}{\includegraphics{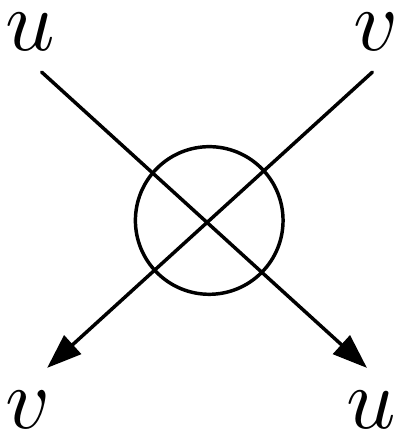}}} \hspace{0.5cm}
\subfigure[]{\scalebox{0.6}{\includegraphics{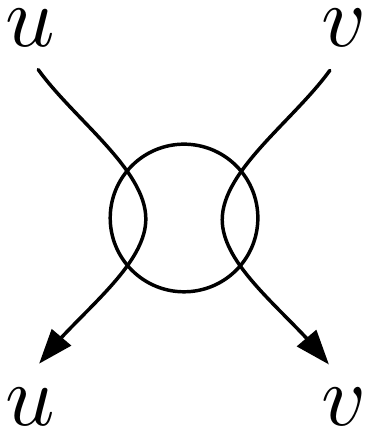}}}
\end{center}
\caption{Schematics of ballistic collisions implemented with gliders.}
\label{particles}
\end{figure}

Ballistic collisions of gliders, or particles, permit to represent functions of two arguments (general signal-interaction scheme is conceptualized in Fig.~\ref{particles}a), as follows~\cite{kn:Toff02}:

{\small
\begin{enumerate}
\item $f(u,v)$ is a product of one collision (Fig.~\ref{particles}b);
\item $f_i(u,v) \mapsto (u,v)$ identity (Fig.~\ref{particles}c);
\item $f_r(u,v) \mapsto (v,u)$ reflection (Fig.~\ref{particles}d);
\end{enumerate}
}
\noindent where Fig.~\ref{particles}a represents a collision not preserving identity of gliders $u$ and $v$; Fig.~\ref{particles}b shows a collision where identities of input gliders are preserved.

\begin{figure}
\begin{center}
\subfigure[]{\scalebox{0.263}{\includegraphics{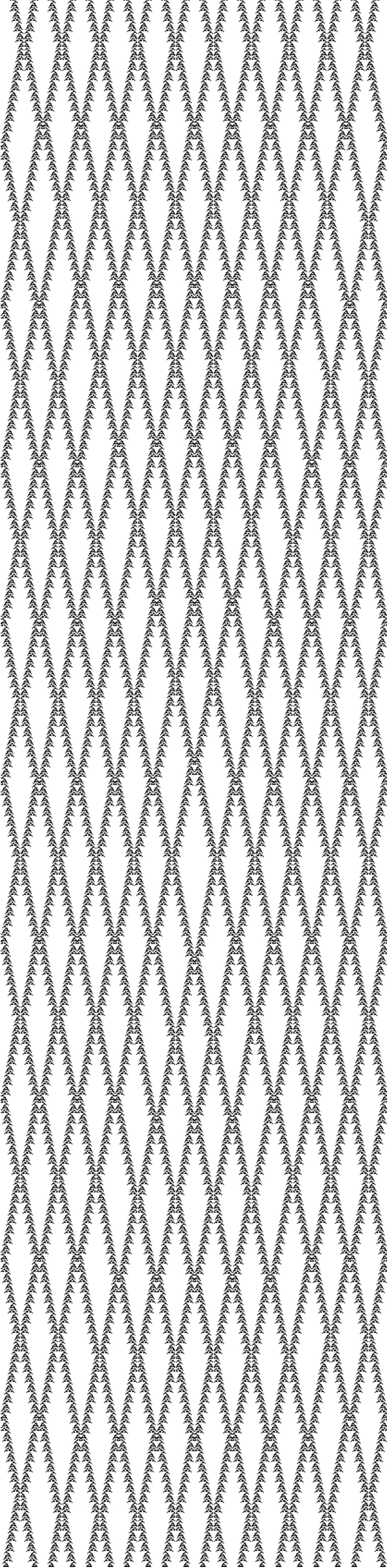}}} \hspace{1.8cm}
\subfigure[]{\scalebox{0.263}{\includegraphics{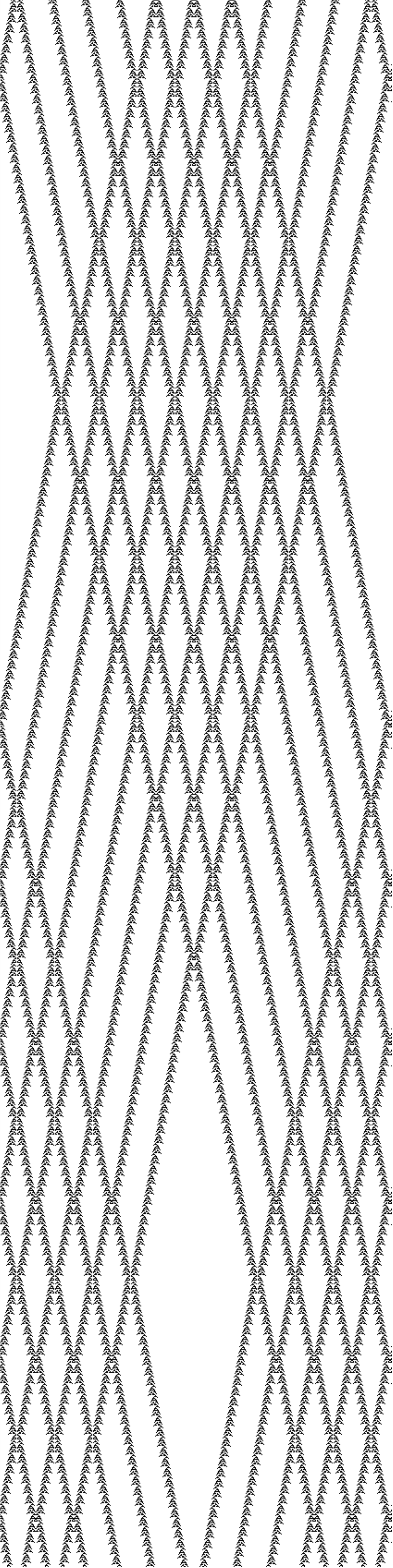}}} \end{center}
\caption{Implementation of ballistic collisions in ECAM rule $\phi_{R22maj:4}$. 
(a) 20 gliders are synchronised to simulate the function identity $f_i(u,v) \mapsto (u,v)$, 
(b) 18 gliders preserve their identity during collisions. ECAM is a ring of 418 cells. It evolved in 1697 time steps.}
\label{ballistic-identity}
\end{figure}

Below we show that ECAM rule $\phi_{R22maj:4}$ reproduces ballistic collisions. Figure~\ref{ballistic-identity} demonstrates identity collisions $f_i(u,v) \mapsto (u,v)$. The collisions between gliders are illustrated in (Fig.~\ref{ballistic-identity}). The collisions are of soliton nature~\cite{kn:JSS01, kn:Steig16}. A carry-ripple adder  \cite{kn:SKW88} can be implemented in ECAM Rule $\phi_{R22maj:4}$ by solitonic reactions with pairs of gliders and identity function (Fig.~\ref{ballistic-identity}a). 

\begin{figure}
\begin{center}
\subfigure[]{\scalebox{0.263}{\includegraphics{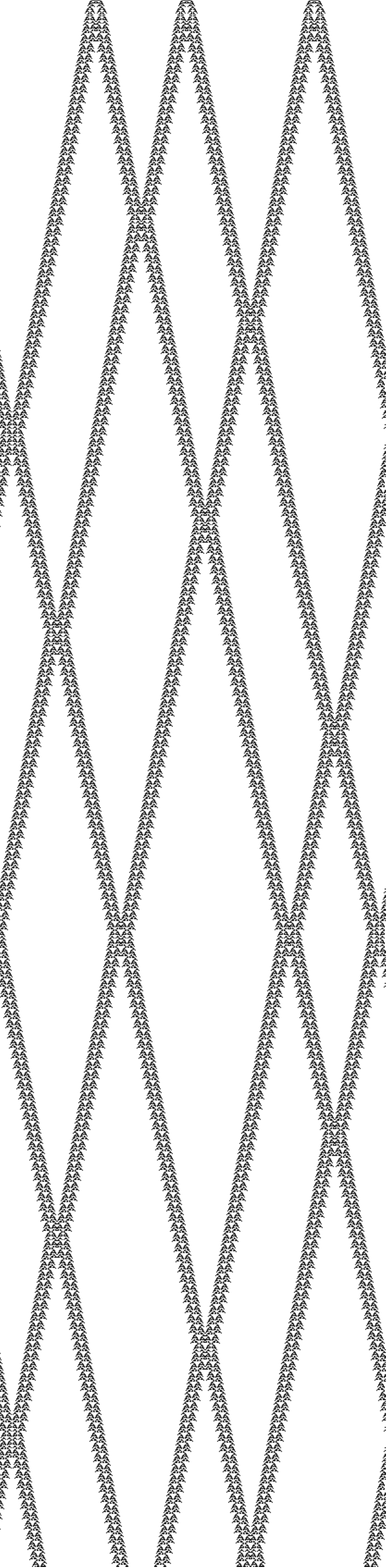}}} \hspace{1.8cm}
\subfigure[]{\scalebox{0.263}{\includegraphics{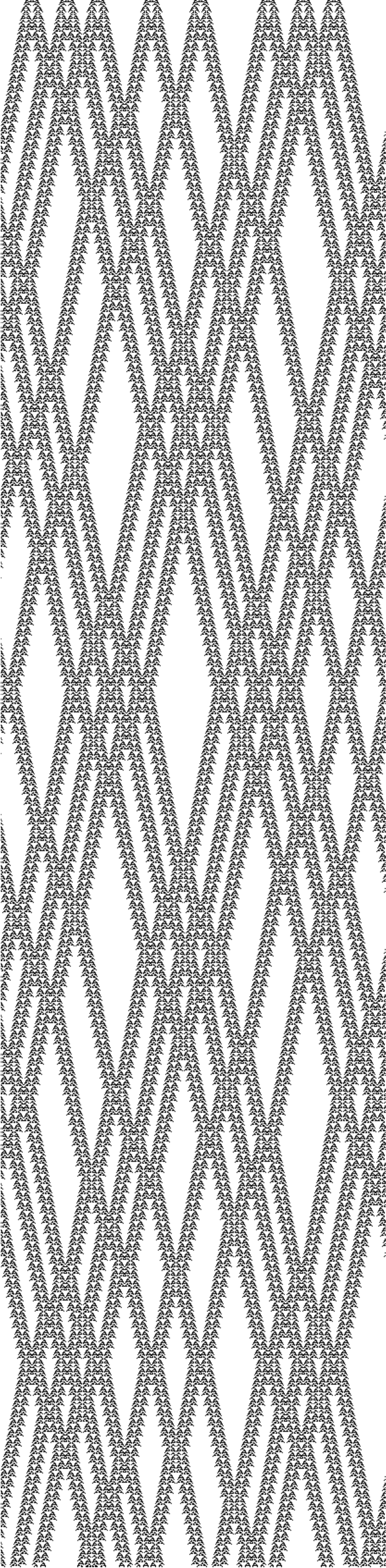}}} \end{center}
\caption{Implementation of ballistic collisions in ECAM rule $\phi_{R22maj:4}$. 
(a) Pairs of six gliders are synchronised to simulate the reflections or elastic collisions $f_r(u,v) \mapsto (v,u)$, (b) 16 pairs of gliders preserve their identity during collisions. ECAM is a ring of 418 cells. It evolved in 1697 time steps.}
\label{ballistic-elastic}
\end{figure}

Figure~\ref{ballistic-elastic} illustrates elastic collisions $f_r(u,v) \mapsto (v,u)$ with a pair of gliders. The pairs of gliders reflect in the same manner in all collisions (Fig.~\ref{ballistic-elastic}a). These elastic collisions are robust to phase changes and initial positions of gliders (Fig.~\ref{ballistic-elastic}b).

\begin{figure}[th]
\begin{center}
\subfigure[]{\scalebox{0.7}{\includegraphics{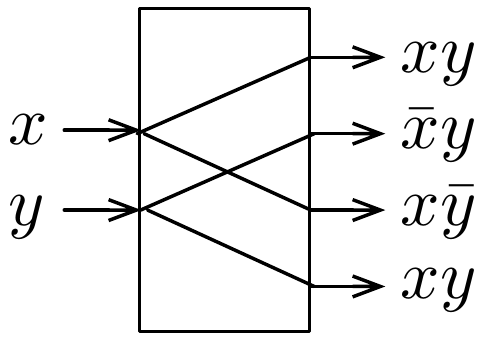}}} \hspace{0.7cm}
\subfigure[]{\scalebox{0.7}{\includegraphics{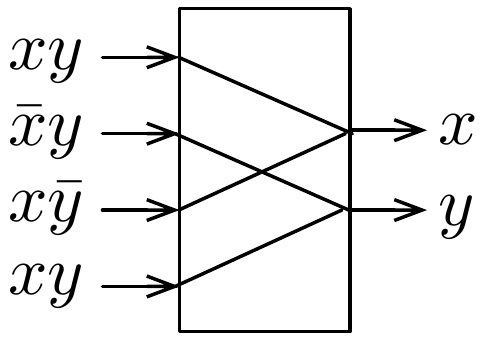}}} \hspace{0.7cm}
\subfigure[]{\scalebox{0.7}{\includegraphics{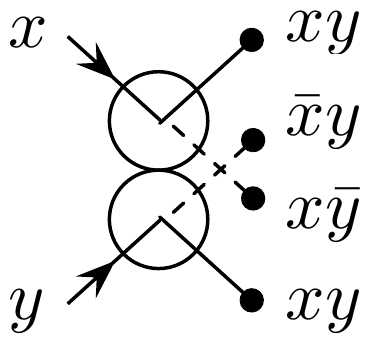}}}
\end{center}
\caption{In ballistic collisions we can implement an (a) interaction gate, (b) its inverse, and (c) its billiard ball model realisation.}
\label{ballistic-2}
\end{figure}

The ballistic interaction gate (Fig.~\ref{ballistic-2}a) is invertible (Fig.~\ref{ballistic-2}b)~\cite{kn:Benn73}. Therefore inverse and elastic collisions in $\phi_{R22maj:4}$ may model the interaction gate: gliders $g_R$ and $g_L$ are equivalent to balls. To allow gliders to return to the original input locations we may constrain them with boundary conditions (Figs.~\ref{ballistic-identity} and~\ref{ballistic-elastic}) or route them with mirrors.

\subsection{Fredkin gate}

\begin{figure}[th]
\begin{center}
\subfigure[]{\scalebox{0.42}{\includegraphics{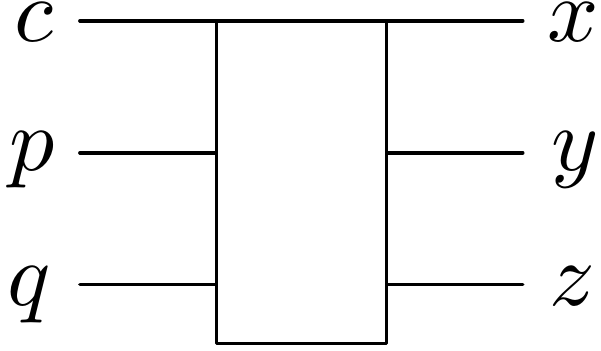}}} \hspace{0.8cm}
\subfigure[]{\scalebox{0.42}{\includegraphics{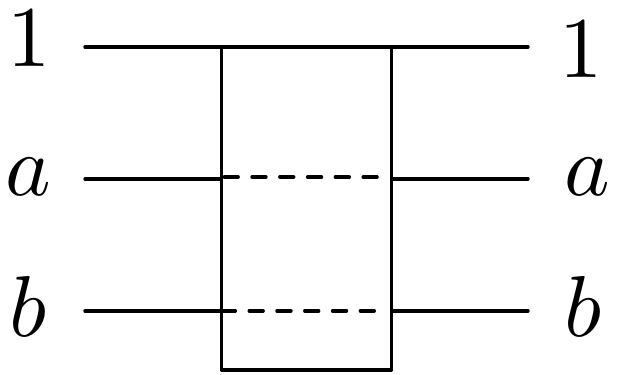}}} \hspace{0.8cm}
\subfigure[]{\scalebox{0.42}{\includegraphics{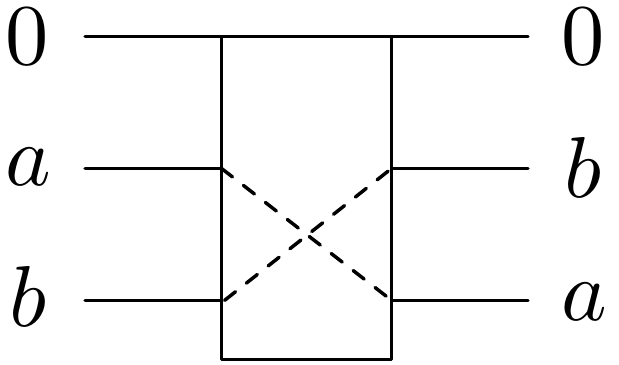}}}
\end{center}
\caption{Fredkin gate. 
(a) Scheme of the gate, 
(b) operation for control value 1,  
(c) operation for control value 0.}
\label{FredkinGate-1}
\end{figure}

A {\it conservative logic gate} is a Boolean function that is invertible and preserve signals \cite{kn:FT82}. Fredkin gate is a classical conservative logic gate.  The gate realises the transformation $(c,p,q) \mapsto (c, cp + \bar{c}q, cq + \bar{c}p)$, where $(c,p,q) \in \{0,1\}^3$. Schematic functioning of Fredkin gate is shown in Fig.~\ref{FredkinGate-1} and the truth table is in Table~\ref{FredkinGateTable}.

\begin{table}[th]
\begin{center}
\begin{tabular}{ccccccc}
$c$ & $p$ & $q$ & & $x$ & $y$ & $z$ \\
0 & 0 & 0 & & 0 & 0 & 0 \\
0 & 0 & 1 & & 0 & 1 & 0 \\
0 & 1 & 0 & & 0 & 0 & 1 \\
0 & 1 & 1 & $\rightarrow$ & 0 & 1 & 1 \\
1 & 0 & 0 & & 1 & 0 & 0 \\
1 & 0 & 1 & & 1 & 0 & 1 \\
1 & 1 & 0 & & 1 & 1 & 0 \\
1 & 1 & 1 & & 1 & 1 & 1 \\
\end{tabular}
\caption{Truth table of Fredkin gate.}
\label{FredkinGateTable}
\end{center}
\end{table}

\begin{figure}
\begin{center}
\subfigure[]{\scalebox{0.45}{\includegraphics{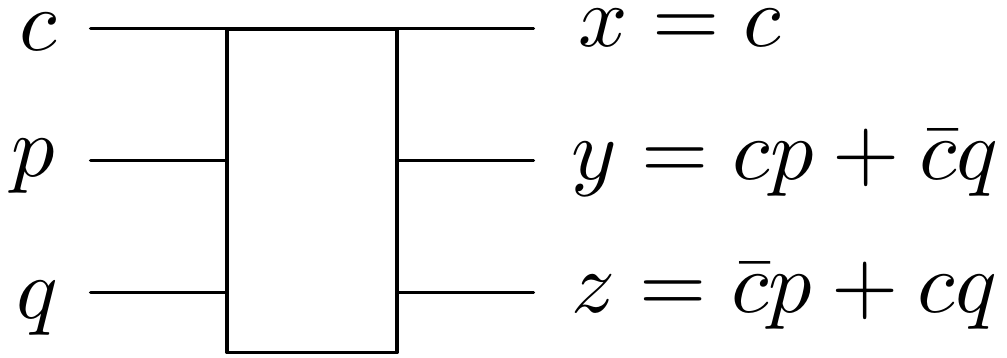}}} \hspace{4.5cm}
\subfigure[]{\scalebox{0.45}{\includegraphics{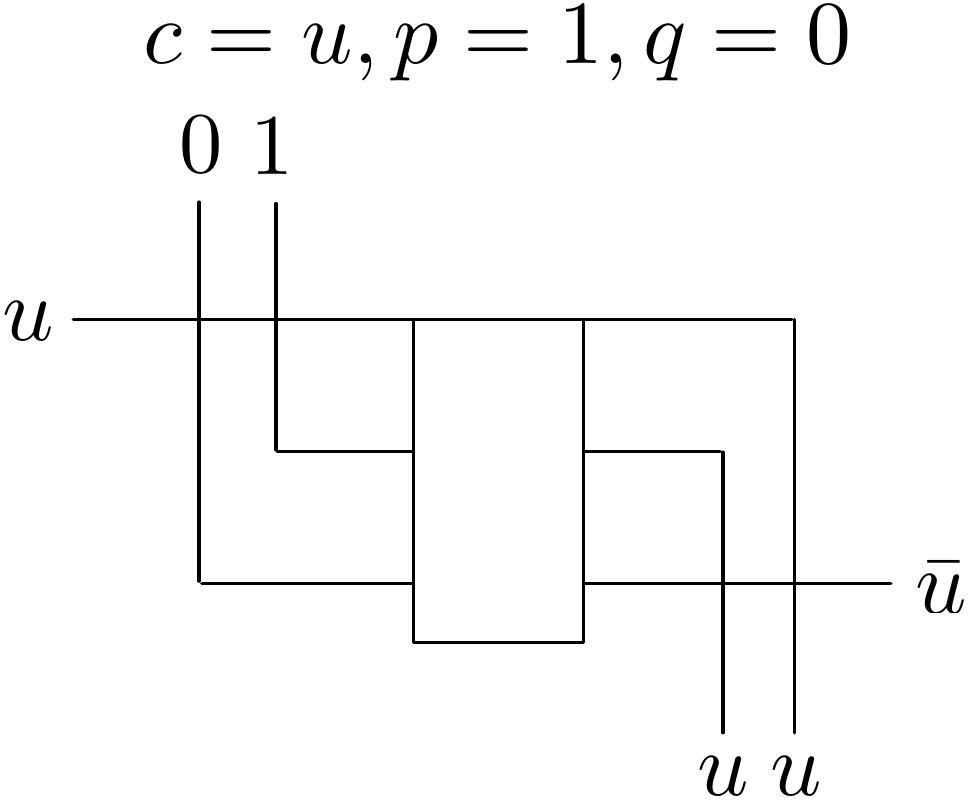}}} \hspace{1.5cm}
\subfigure[]{\scalebox{0.45}{\includegraphics{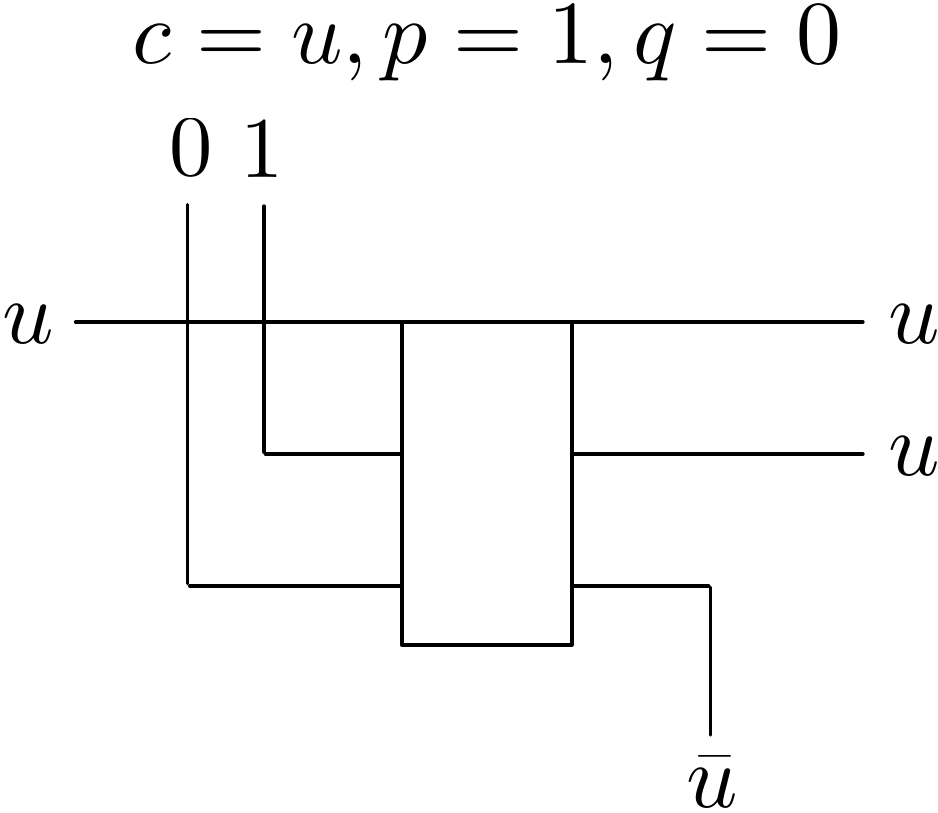}}} \hspace{1.5cm}
\subfigure[]{\scalebox{0.45}{\includegraphics{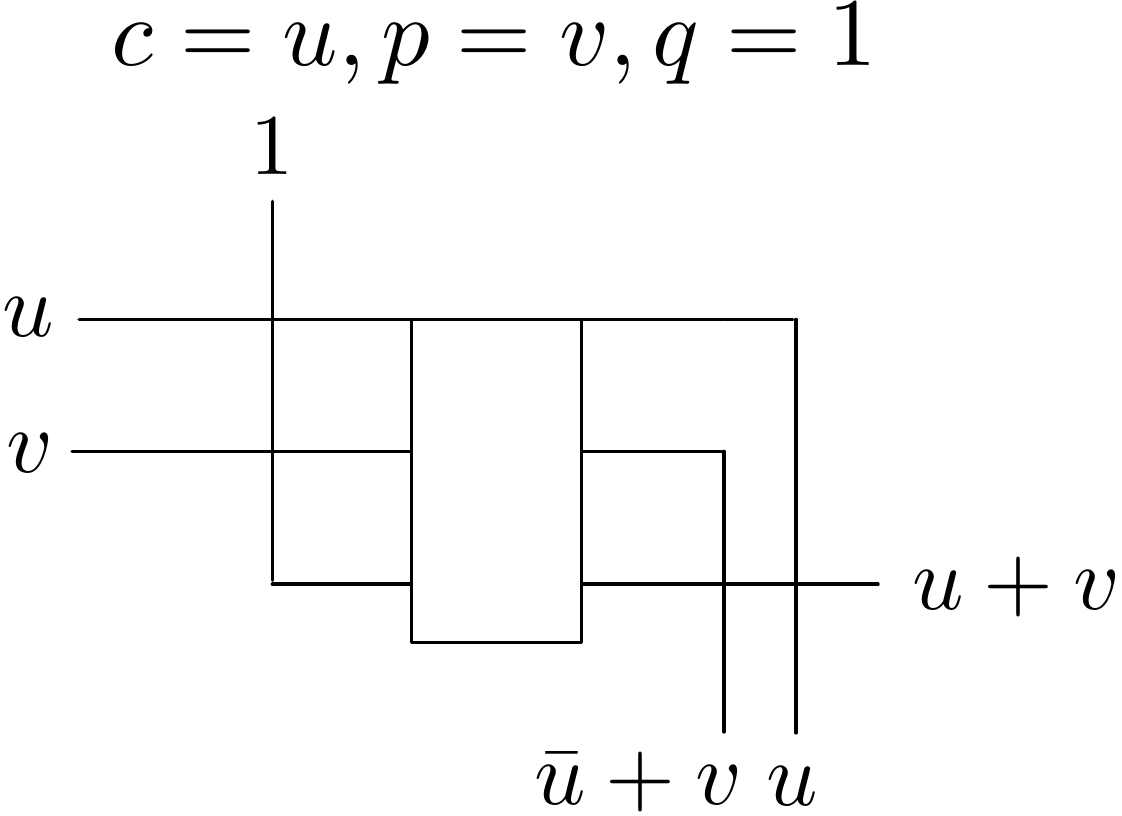}}} \hspace{1.5cm}
\subfigure[]{\scalebox{0.45}{\includegraphics{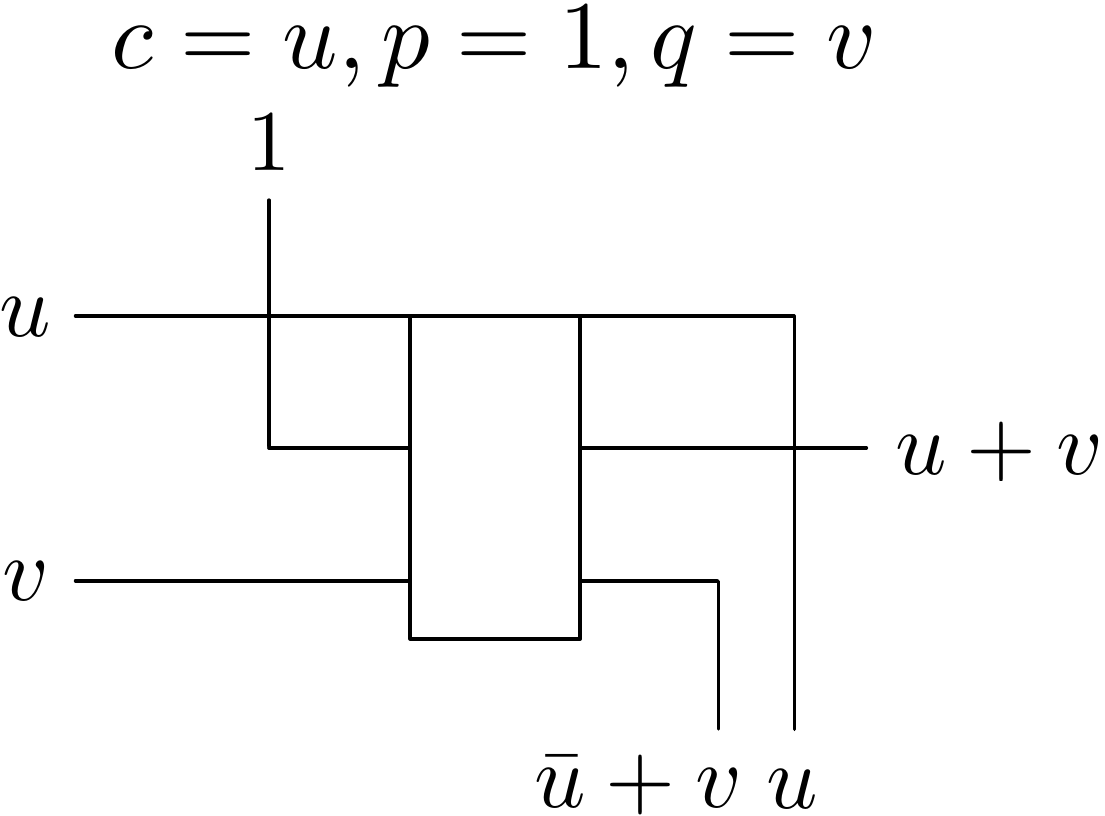}}} \hspace{1.5cm}
\subfigure[]{\scalebox{0.45}{\includegraphics{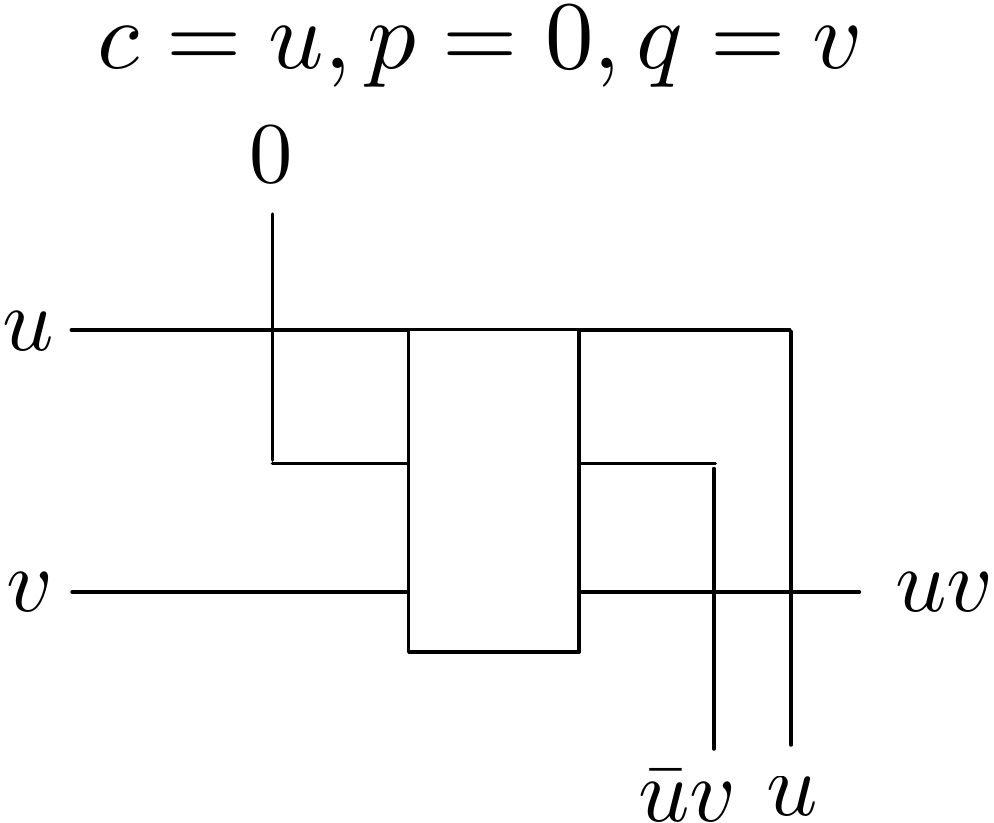}}} \hspace{1.5cm}
\subfigure[]{\scalebox{0.45}{\includegraphics{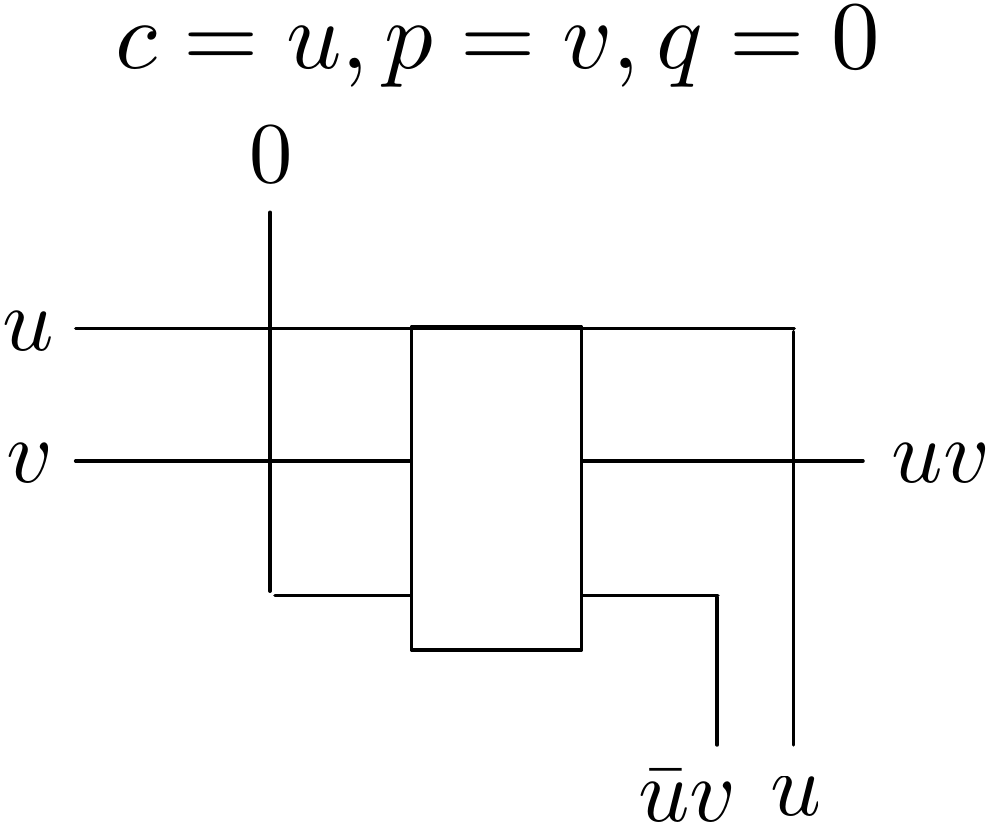}}}
\end{center}
\caption{Realisation of Boolean functions using Fredkin gate. 
(a) Fredkin gate, 
(b) {\sc not} gate,
(c) {\sc fanout} gate, 
(d-e) {\sc or} gate, and 
(f-g) {\sc and} gate \cite{kn:FT82}.}
\label{FredkinGate-2}
\end{figure}

Fredkin gate is universal because one can implement a functionally complete set of logical functions with this gate (Fig.~\ref{FredkinGate-2}). Other gates implemented with Fredkin gate are shown below:

\begin{itemize}
\item $c=u$, $p=v$, $q=1$ (or $c=u$, $p=1$, $q=v$) yields the {\sc implies} gate $y= u \rightarrow v$ $(z = u \rightarrow v)$.
\item $c=u$, $p=0$, $q=v$ (or $c=u$, $p=v$, $q=0$) yields the {\sc not implies} gate $y = \overline{(v \rightarrow u)}$ $(z= \overline{(v \rightarrow u)})$ \cite{kn:FT82}.
\end{itemize}

\subsection{Basic collisions}

To simulate a Fredkin gate in ECAM rule $\phi_{R22maj:4}$, we utilise  a set of collisions to preserve the reactions and the persistence of data, these basic collisions have a specific task and actions which are specified as follows.

\begin{itemize}
\item {\it Mirror} reflects a glider,  the mirror should be deleted, not to become an obstacle for other signals.
\item {\it Doubler} splits a signal into two signals.
\item {\it Soliton} crosses two gliders preserving its identity but might change its phase.
\item {\it Splitter} separates two gliders into gliders travelling in opposite directions.
\item {\it Flag} is a glider that is generated depending of an input value given.
\item {\it Displacer} moves a glider forward.
\end{itemize}

Fredkin gates implemented in non-invertible systems were proposed and simulated by Adamatzky in a non-linear medium --- Oregonator model of Belousov-Zhabotinsky \cite{kn:Ada17}.

\begin{figure}[th]
\centerline{\includegraphics[width=4.7in]{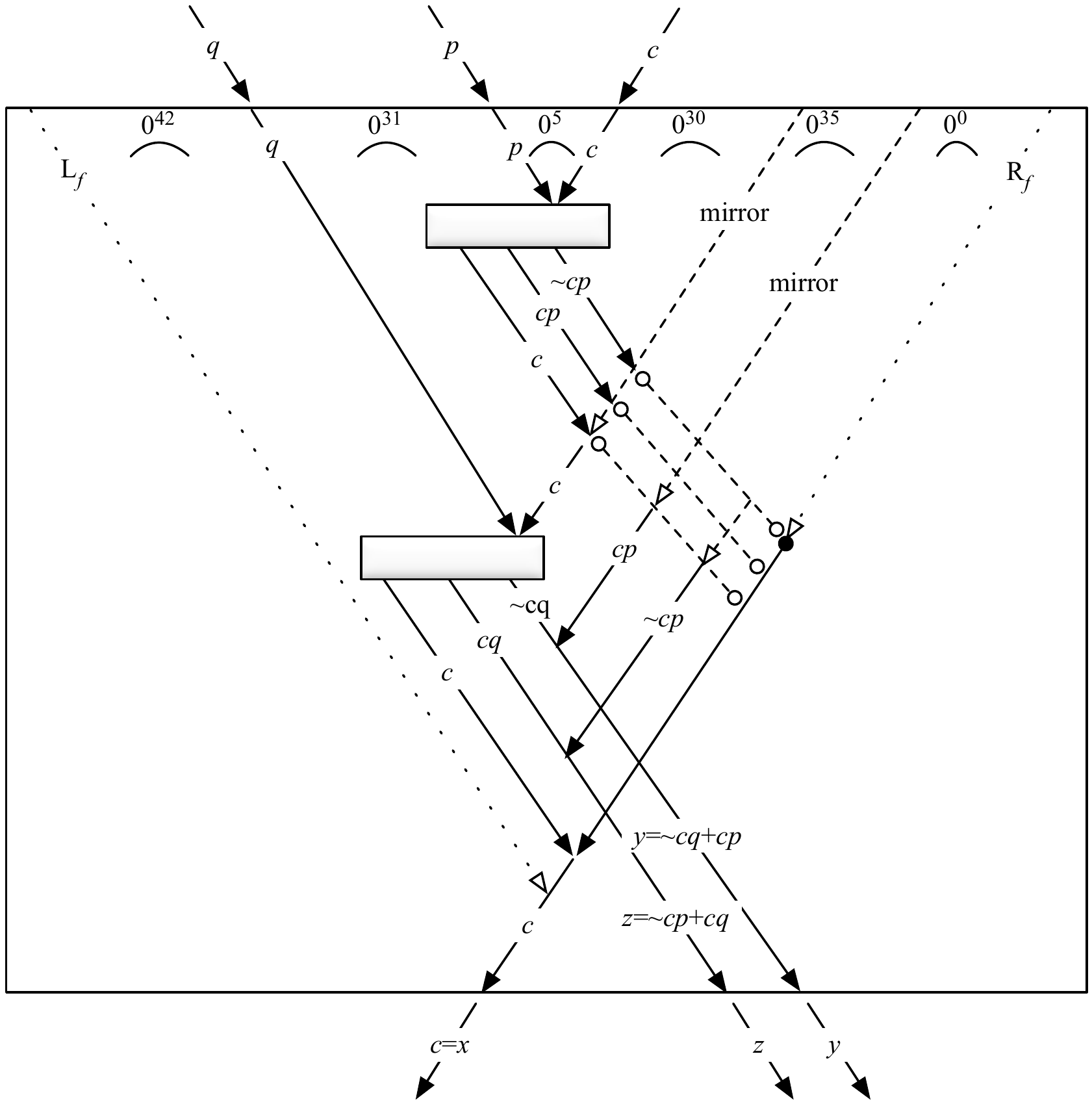}}
\caption{A scheme of Fredkin gate implementation in ECAM rule $\phi_{R22maj:4}$ via glider collisions.}
\label{diagCollFredkinGate}
\end{figure}

\subsection{Fredkin gates in one dimension}

Figure~\ref{diagCollFredkinGate} displays the schematic diagram proposed to simulate Fredkin gates in ECAM rule $\phi_{R22maj:4}$. There are three inputs $(c,p,q)$ and three outputs $(x,y,z)$. During the computation auxiliary gliders are generated. They are deleted before reaching outputs. Gliders travel in two directions in a 1D chain of cells, they can be reused only when crossing periodic boundaries or via combined collisions. We use two gliders as flags, travelling from the left `$L_f$' and from the right `$R_f$'. Flags are activated depending on initial values for $c$ or $q$ as follows

\begin{itemize}
\item[] If $c=0$ then $R_f=1$,
\item[] If $q=0$ then $L_f=1$,
\item[] In any other case $L$ and $R = 0$.
\end{itemize}

Mirrors $M$ are defined as follows:

\begin{itemize}
\item[] If $c=p=q=1$ then $M=2$,
\item[] If $c=p=1$ then $M=2$,
\item[] If $c=q=1$ then $M=1$,
\item[] If $p=1$ then $M=1$,
\item[] In any other case $M=0$.
\end{itemize}

Distances between gliders are fixed as positive integers determined for a number of cells in the state `0', as $\stackrel{\text{\tiny $0^n$}}{\frown}$ $\forall$ $n \geq 0$. During the computation we split gliders  when two gliders travel together, the split gliders travel in opposite directions. We use two gliders as mirrors to change the direction of an argument movement. We use displacer  to move an output glider and adjust its distance with respect to other. 

\begin{table}[th]
\begin{center}
\begin{tabular}{ c|cccc }
$cpq$ & $xyz$ & $L_f$ & $R_f$ & M \\
\hline
111 & 111 & 0 & 0 & 2 \\
110 & 110 & 1 & 0 & 2 \\
101 & 101 & 0 & 0 & 1 \\
100 & 100 & 1 & 0 & 0 \\
011 & 011 & 0 & 1 & 0 \\
010 & 001 & 1 & 1 & 1 \\
001 & 010 & 0 & 1 & 0 \\
\end{tabular}
\caption{Following the scheme in Fig.~\ref{diagCollFredkinGate} we specify a sequence of collisions that are controlled  with  flag gliders ($L_f$, $R_f$) and mirrors (M) glider.}
\label{tableElementsCollions}
\end{center}
\end{table}

Table~\ref{tableElementsCollions} shows values of inputs, outputs, flags, and mirrors. First column represents {\sc inputs}, the second column {\sc outputs}, third column are values of a flag activated in the left side, fourth column are values of the flag activated in the right side, and the last column shows the number of necessary mirrors.

\begin{figure}
\begin{center}
\subfigure[]{\scalebox{0.3}{\includegraphics{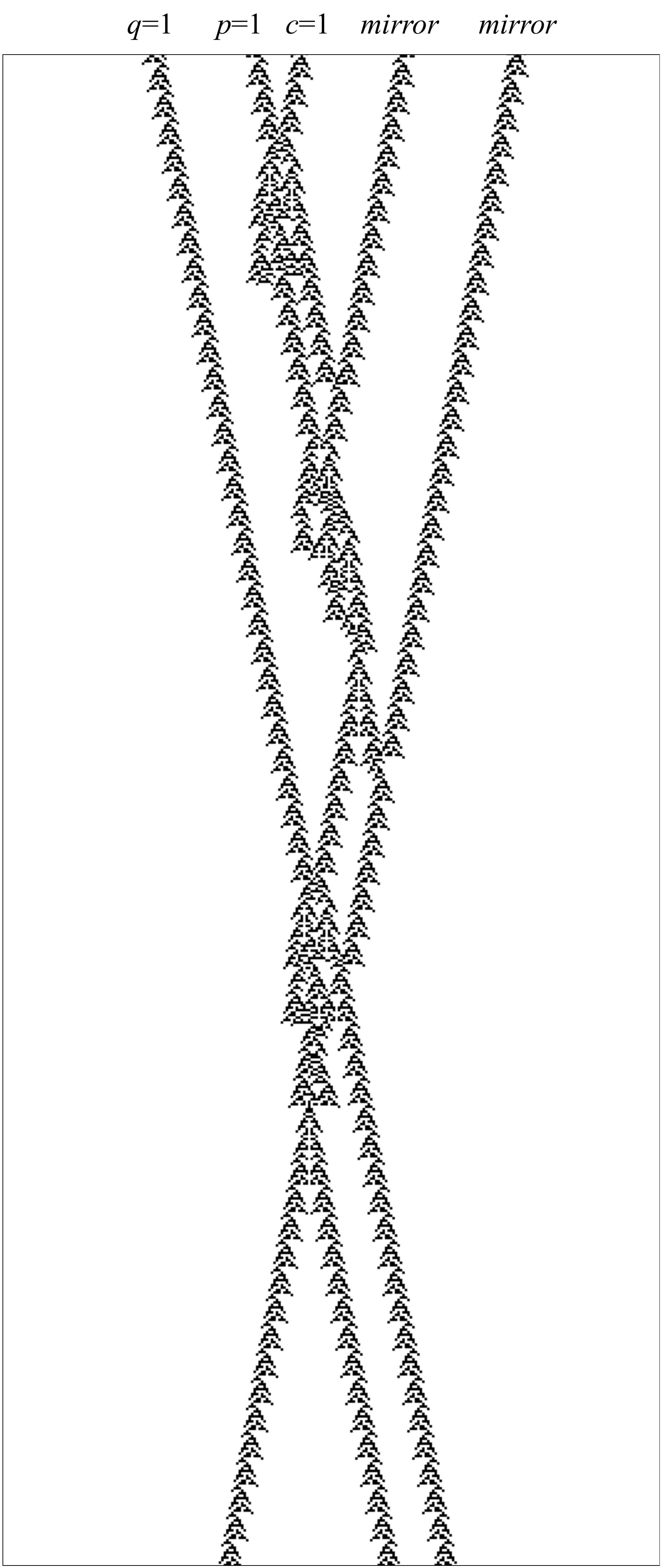}}} \hspace{0.5cm}
\subfigure[]{\scalebox{0.3}{\includegraphics{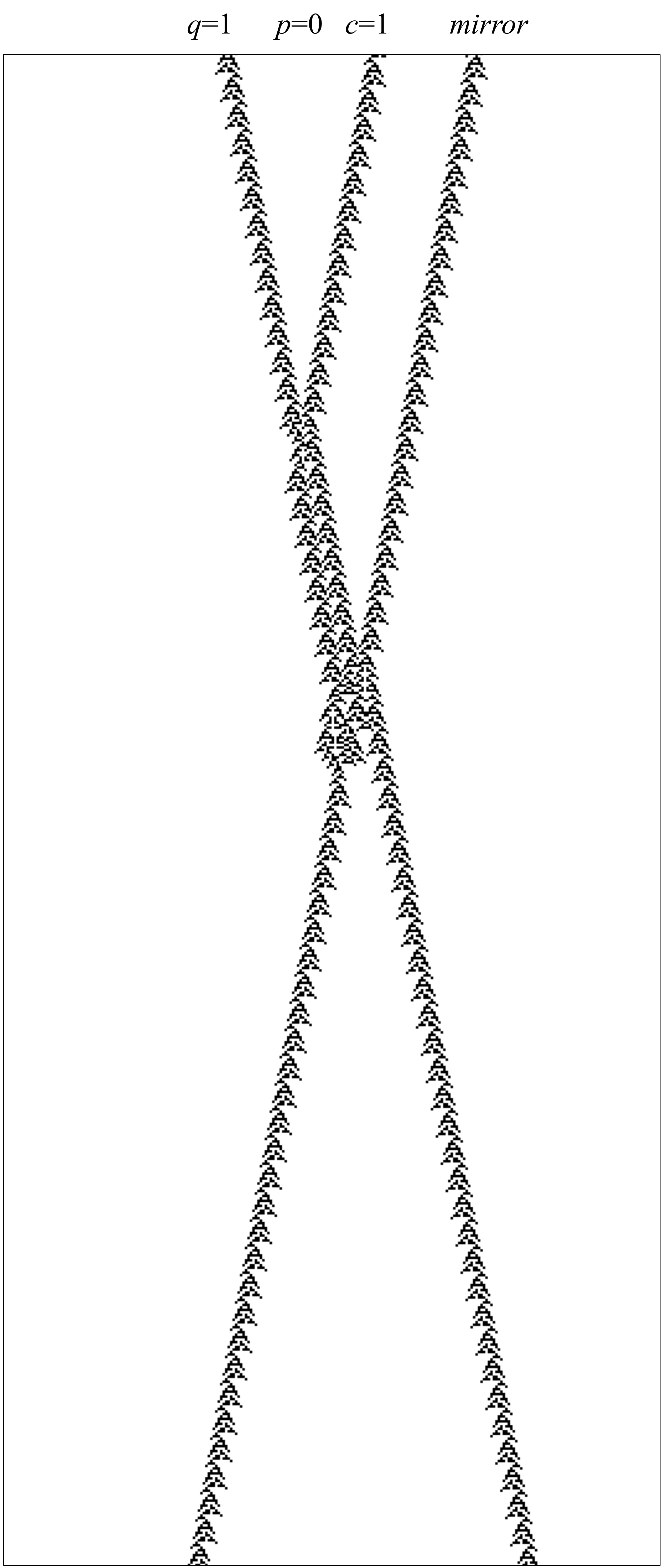}}}
\end{center}
\caption{Fredkin gate in ECAM rule $\phi_{R22maj:4}$.  
(a) {\sc INPUT} $c=1$, $p=1$, $q=1$, {\sc OUTPUT} $x=1$, $y=1$, $z=1$,
(b) {\sc INPUT} $c=1$, $p=0$, $q=1$, {\sc OUTPUT} $x=1$, $y=0$, $z=1$.}
\label{FredkinGateab}
\end{figure}

\begin{figure}
\begin{center}
\subfigure[]{\scalebox{0.29}{\includegraphics{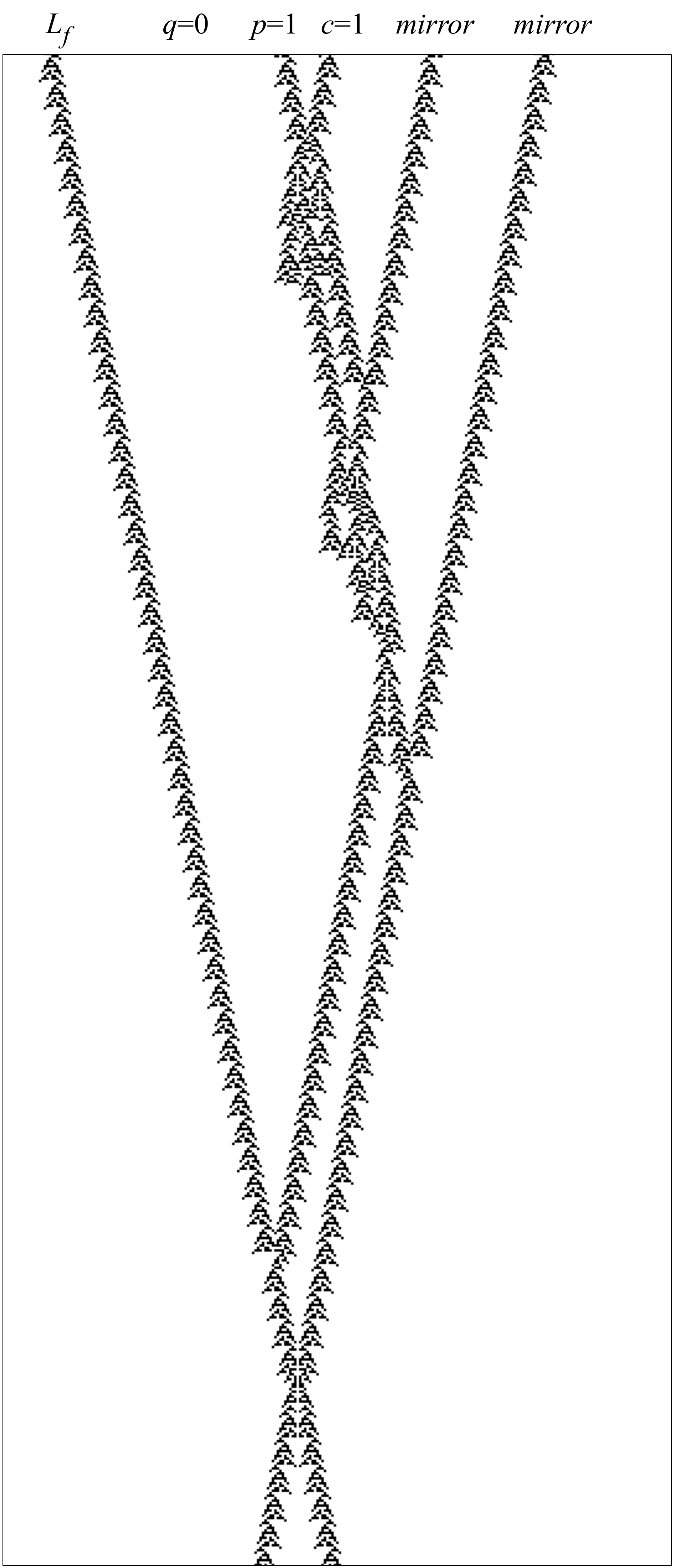}}} \hspace{0.5cm}
\subfigure[]{\scalebox{0.29}{\includegraphics{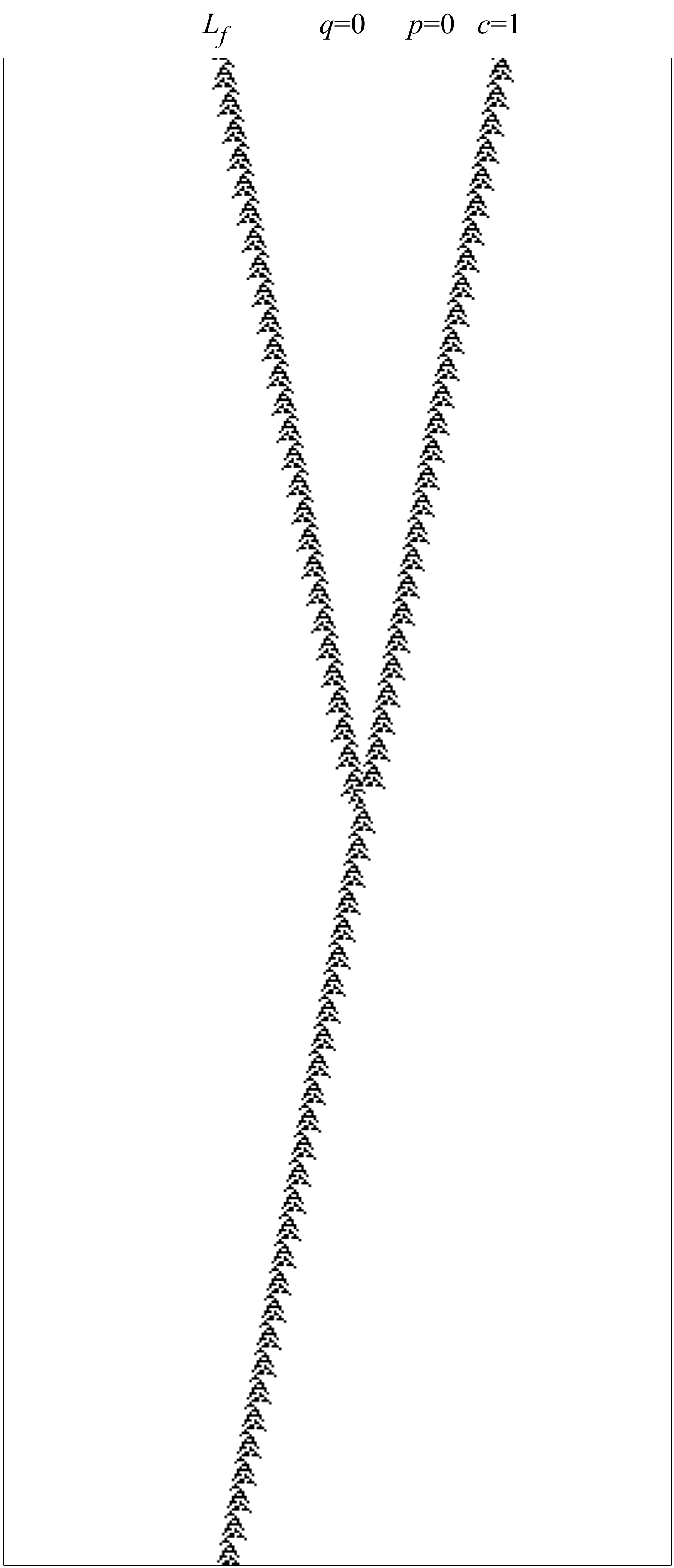}}}
\end{center}
\caption{Fredkin gate in ECAM rule $\phi_{R22maj:4}$. 
(a) {\sc INPUT} $c=1$, $p=1$, $q=0$, {\sc OUTPUT} $x=1$, $y=1$, $z=0$, 
(b) {\sc INPUT} $c=1$, $p=0$, $q=0$, {\sc OUTPUT} $x=1$, $y=0$, $z=0$.}
\label{FredkinGatecd}
\end{figure}

\begin{figure}
\begin{center}
\subfigure[]{\scalebox{0.28}{\includegraphics{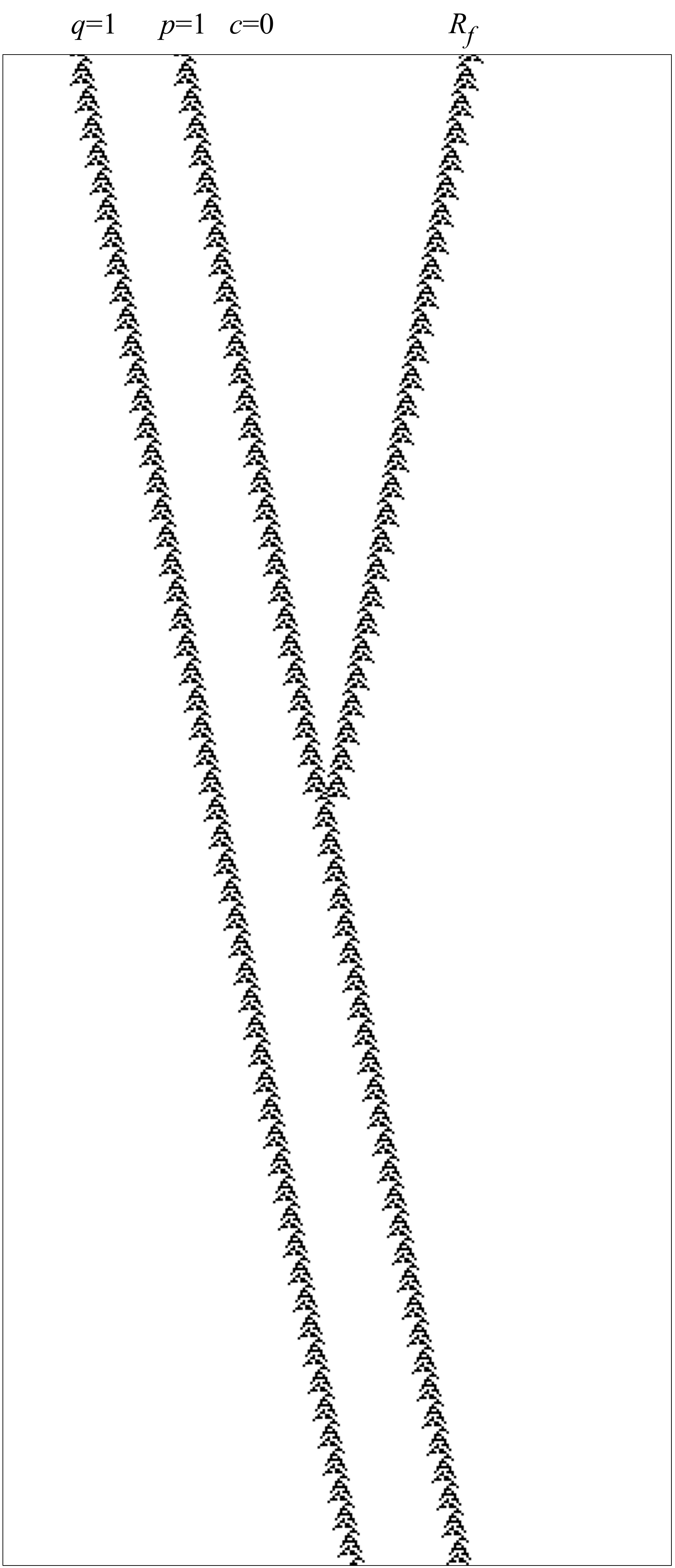}}} \hspace{0.5cm}
\subfigure[]{\scalebox{0.28}{\includegraphics{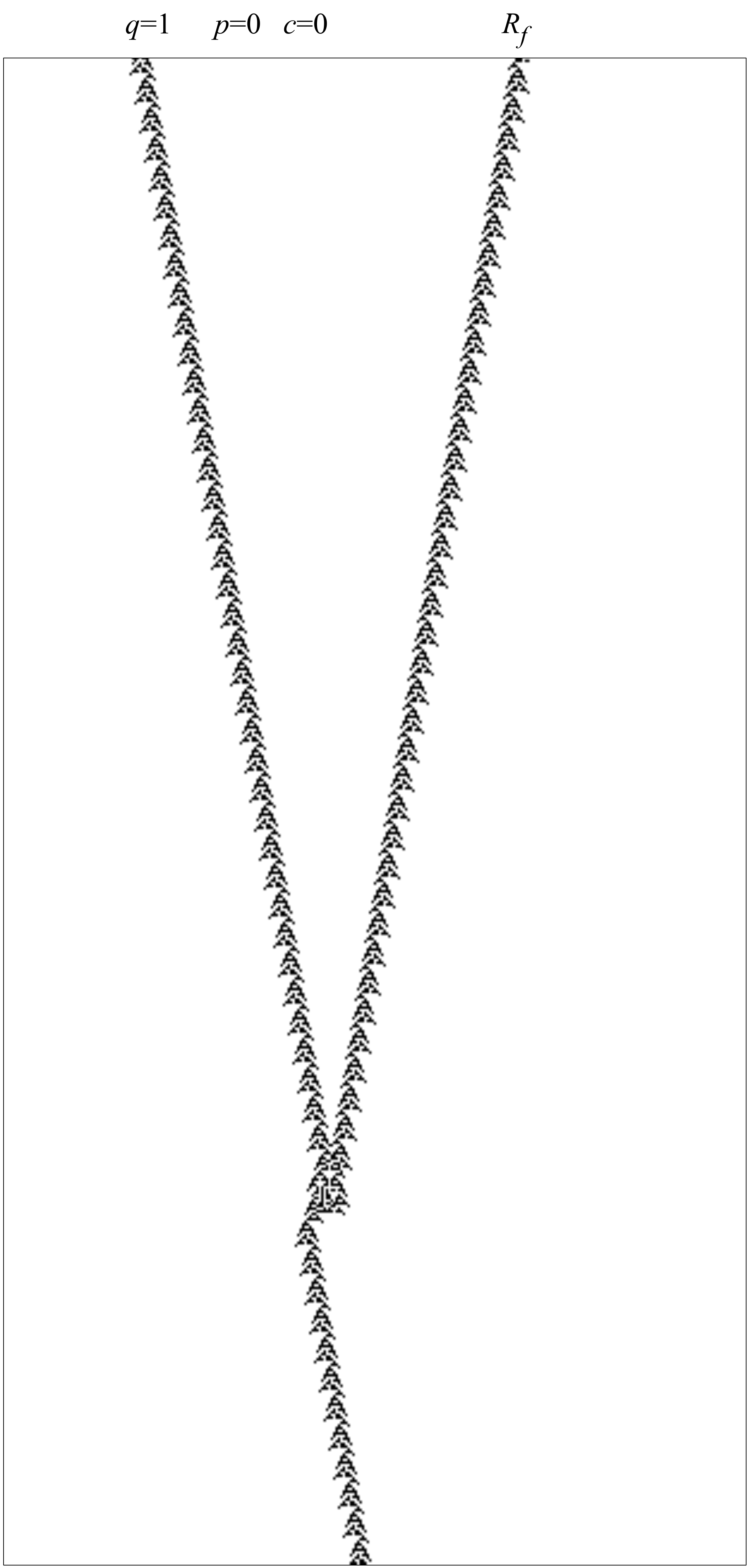}}}
\end{center}
\caption{Fredkin gate in ECAM rule $\phi_{R22maj:4}$. 
(a) {\sc INPUT} $c=0$, $p=1$, $q=1$, {\sc OUTPUT} $x=0$, $y=1$, $z=1$, 
(b) {\sc INPUT} $c=0$, $p=0$, $q=1$, {\sc OUTPUT} $x=0$, $y=1$, $z=0$.}
\label{FredkinGateef}
\end{figure}

\begin{figure}
\centerline{\includegraphics[width=2in]{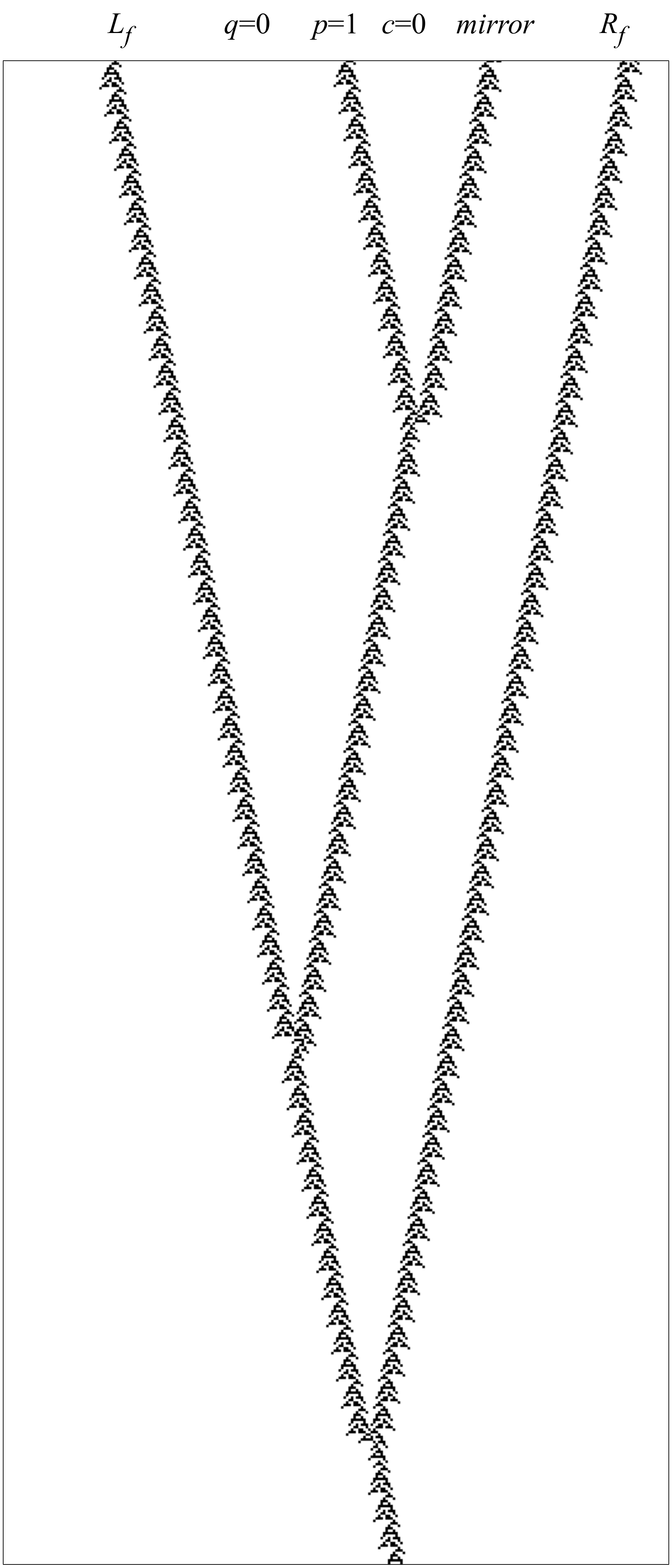}}
\caption{Fredkin gate in ECAM rule $\phi_{R22maj:4}$. 
{\sc INPUT} $c=0$, $p=1$, $q=0$, {\sc OUTPUT} $x=0$, $y=0$, $z=1$.}
\label{FredkinGateg}
\end{figure}

Space-time configurations of ECAM rule $\phi_{R22maj:4}$ implementing Fredkin gate for all non-zero combinations of inputs are shown in Figs.~\ref{FredkinGateab}--\ref{FredkinGateg}. 

\begin{figure}
\centerline{\includegraphics[width=2.7in]{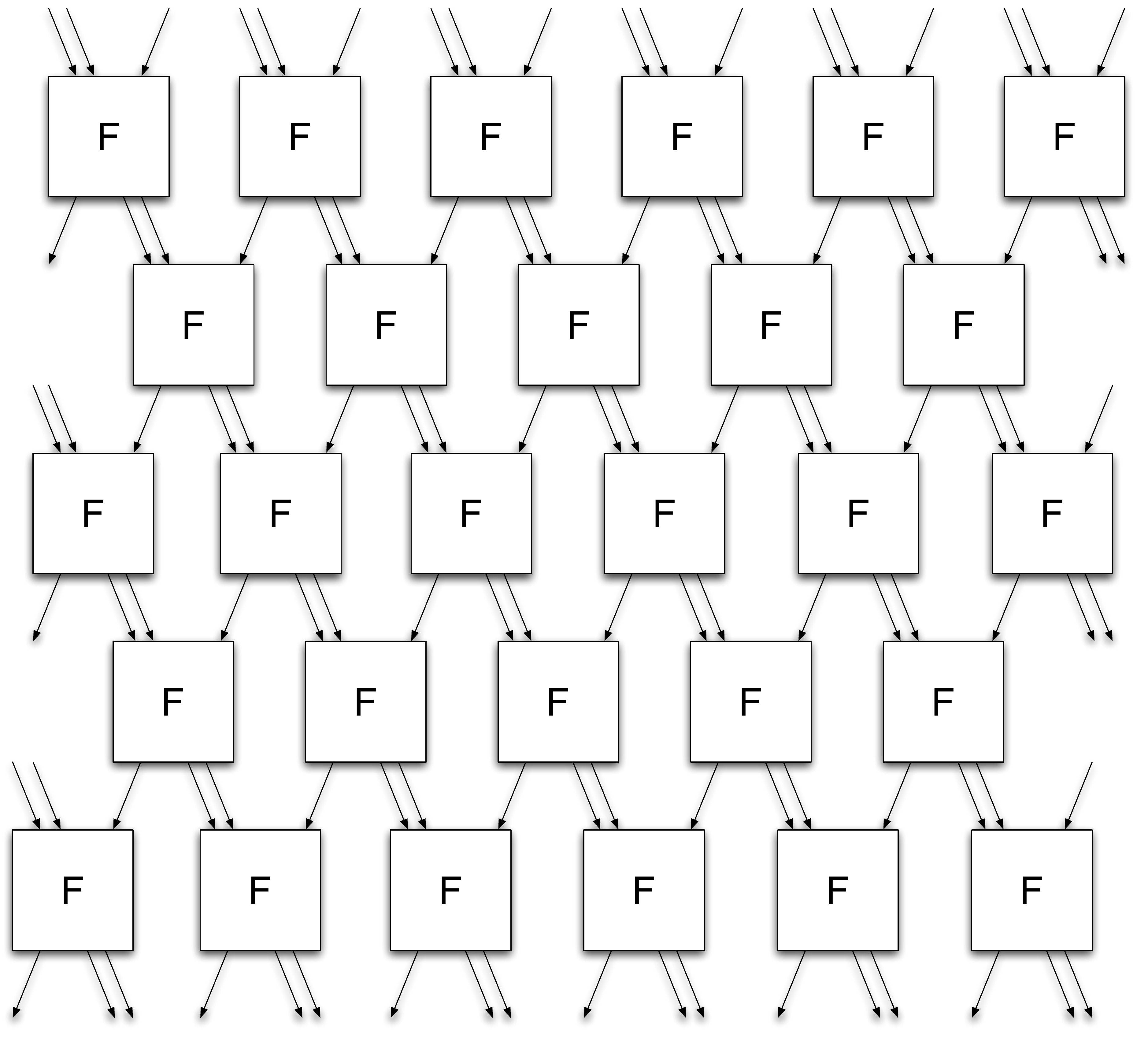}}
\caption{Cascaded Fredkin gates (It is a modification of Fredkin array proposed ~\cite{kn:Marg02}).}
\label{FredkinCircuit}
\end{figure}

\section{Discussion}

We demonstrated how to implement a functionally complete set of Boolean functions in a one-dimensional cellular automaton with binary cell states, three-cell neighbourhood and memory depth four. We also shown that Fredkin gate can be realised the this automaton. Let us compare complexity of our designs with previously published models of universal one-dimensional cellular automata.

CA simulating Turing machine, proposed by Smith III in 1971~\cite{kn:Smi71}, satisfied the condition: a number of cell-states multiplied by a neighbourhood size equals $\eta=36$. A universal 1D CA designed by Albert and Culik II~\cite{kn:AC87} has 14 states and totalistic cell-state transition rule, that is for their automaton $\eta=42$. The Lindgren-Nordahl CA~\cite{kn:LN90} has 7 states, assuming three-cell neighbourhood we have $\eta=21$. ECAM implementation of a Fredkin gate, proposed in present paper, has two cell states, three cell neighbourhood, and memory depth 4, this implies $\eta=24$. Thus, in terms of a neighbourhood size and a number of states our automaton is less efficient than the CA proposed by Lindgren-Nordahl~\cite{kn:LN90}, however we use gliders as signals. Gliders are discrete analogs of solitons, therefore our design, in principle, can be considered as a blueprint for physical implementation, as we will discuss below. Cook's design~\cite{kn:Cook04, kn:Wolf02} of a universal CA via cyclic tag system is the most optimal in terms of neighbourhood size and number of states, $\eta=6$, however the implementation involves 11 gliders and a glider gun. Our design of Fredkin gate utilises just two types of gliders. 

Polymer chains that support propagation of travelling localisations (solitons, kinks, defects) are potential substrates for implementation of glider-based Fredkin gate. There are many potential candidates, here we discuss actin filaments. Actin is a protein presented in all eukaryotic cells in forms of globular actin (G-actin) and filamentous actin (F-actin), see history overview in~\cite{szent2004early}. Filamentous actin is a double helix of F-actin unit chains. In \cite{adamatzky2015actin} we proposed a model of actin polymer filaments as two chains of one-dimensional binary-state semi-totalistic automaton arrays and uncovered rules supporting gliders, discrete analogs of ionic waves propagating in actin filaments~\cite{tuszynski2004ionic}.  

Let us evaluate a physical space-time requirement for implementation of Fredkin gate on actin filaments. Assume a unit of F-actin corresponds to a cell of 1D CA. Maximum diameter of an actin filament is 8~nm~\cite{moore1970three, spudich1972regulation}. An actin filament is composed of overlapping units of F-actin. Thus, diameter of a single unit is c. 4~nm. A glider in our ECAM model occupies 10 cells, that makes a size of a signal in actin filaments 40 nm. Maximum distance between inputs in ECAM Fredkin gate is 200 cells. This makes gate size 880~nm.  

With regards to speed of Fredkin gate realisation, being unaware of exact mechanisms of travelling localisations on actin filaments we can propose speculative estimates. Assume the underpinning mechanism of generating a localisation is an excitation of F-actin molecule (single unit of actin polymer). An  excitation in a molecule takes place when an electron in a ground state absorbs a photon and moves up to a high yet unstable energy level. Later the electron returns to its ground state. When returning to the ground state the electron releases photon which travels with speed  $\times 10^{18}$ \AA\, per second. F-actin molecule (one unit of actin filament) is a polymer chain of at most 3K nodes. Thus the F-actin unit can be spanned  by an excitation in at most $10^{-15}$ sec. That can be adopted as a physical time step equivalent to one step of ECAM evolution. The ECAM Fredkin gate completes its operation on actin filament in 600 time steps, that is at most $10^{-13}$ sec, i.e. 0.1 picosecond of real time. 

Experimental implementation of the Fredkin gate, including cascading of the gates, on actin filaments, or any other soliton-supporting polymer chain makes a very challenging topic of future studies. There methodological approaches to control and monitor attosecond scale molecular dynamics ~\cite{goulielmakis2008single, baltuvska2003attosecond, nabekawa2017probing, ciappina2017attosecond} however it is not clear if they can be applied to actin polymers. The problems to overcome include input of data in acting filament with a single F-actin unit precision, keeping the polymer chain stable and insulated from thermal noise, reading outputs from the polymer chain. However exact experimental implementation remains uncertain.

But, although this can be confined in a physical computing device several limitations should be improved: avoid the use of mirrors and the use of flags. An option is that they could be manipulated into in a ring (or virtual CA collider \cite{kn:MAS11, kn:MAM16}).


\newpage

\appendix

\section{Appendix -- Binary collisions in $\phi_{R22maj:4}$}

\begin{figure}[th]
\centerline{\includegraphics[width=4.4in]{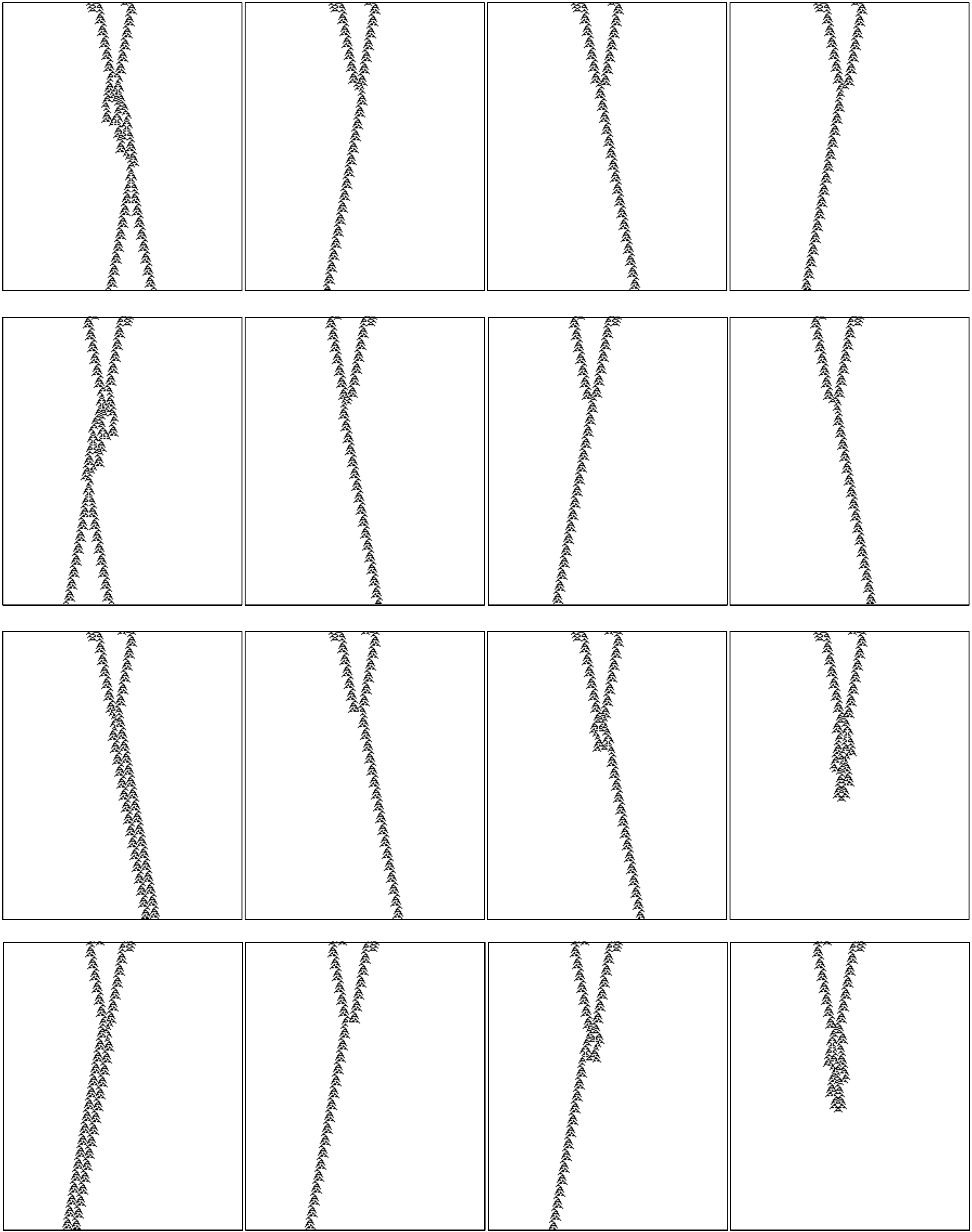}}
\caption{Binary collisions in Rule 22 with memory.}
\label{binCollisions-1}
\end{figure}

\newpage

\begin{figure}[th]
\centerline{\includegraphics[width=4.4in]{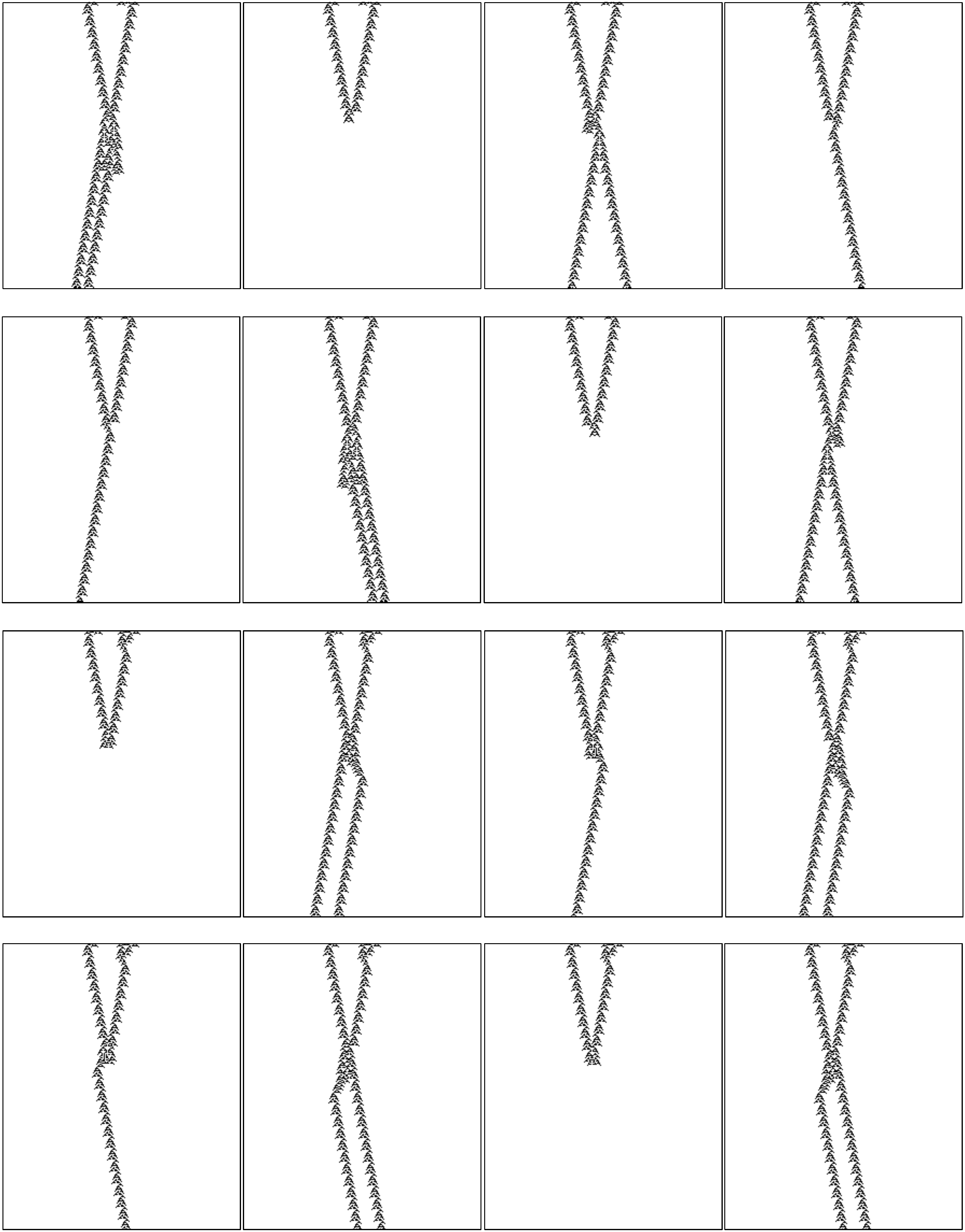}}
\caption{Binary collisions in Rule 22 with memory.}
\label{binCollisions-2}
\end{figure}

\newpage

\begin{figure}[th]
\centerline{\includegraphics[width=4.4in]{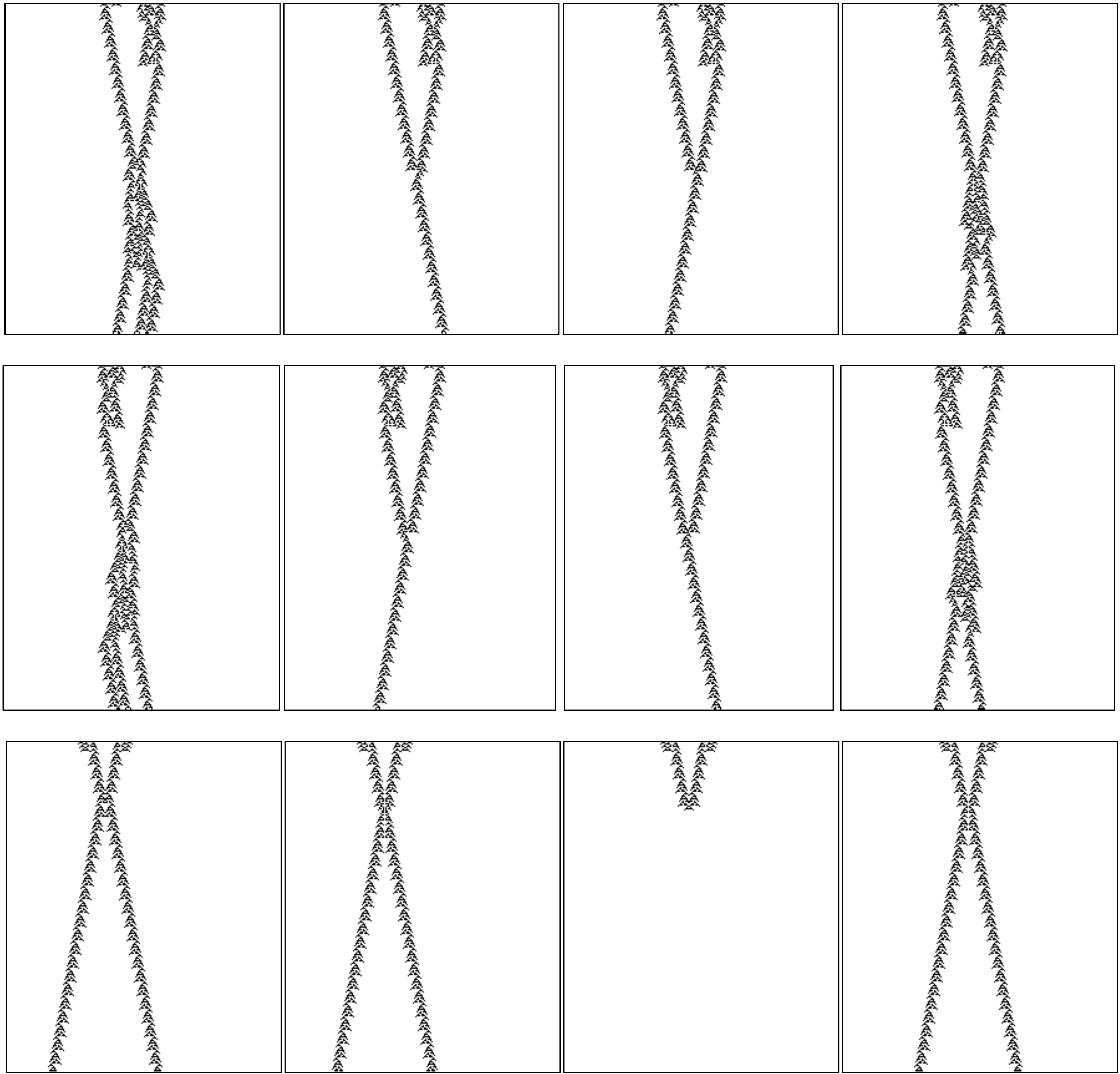}}
\caption{Binary collisions in Rule 22 with memory.}
\label{binCollisions-3}
\end{figure}


\begin{thebibliography}{99}

\bibitem{kn:AC87} Albert, J. \& Culik II, K. (1987) A simple universal cellular automaton and its one-way and totalistic versions, {\em Complex Systems} {\bf 1(1)} 1--16.

\bibitem{kn:Ada01} Adamatzky, A. (2001) {\em Computing in Nonlinear Media and Automata Collectives}, IOP Publishing Ltd. Bristol, UK.

\bibitem{kn:Ada02} Adamatzky, A. (Ed.) (2002) {\em Collision-Based Computing}, Springer.

\bibitem{kn:ACA05} Adamatzky, A., Costello, B.L. \& Asai, T. (2005) {\em Reaction-Diffusion Computers}, Elsevier.

\bibitem{kn:Ada10} Adamatzky, A. (2010) {\em Physarum Machines: Computers from Slime Mould}, World Scientific Series on Nonlinear Science Series A: Volume 74.

\bibitem{adamatzky2015actin} Adamatzky, A. \& Mayne, R. (2015) Actin automata: Phenomenology and localizations, {\em International Journal of Bifurcation and Chaos} {\bf 25(02)} 1550030.

\bibitem{kn:Ada17} Adamatzky, A. (2017) Fredkin and Toffoli gates implemented in oregonator model of Belousov-Zhabotinsky medium, {\em Int. J. Bifurcation and Chaos} {\bf 27(3)} 1750041.

\bibitem{baltuvska2003attosecond} Baltuska, A., Udem, T., Uiberacker, M., Hentschel, M., Goulielmakis, E., Gohle, C., Holzwarth, R., Yakovlev, V.S., Scrinzi, A., Hansch, T.W. \& Krausz, F. (2003) Attosecond control of electronic processes by intense light fields, {\em Nature} {\bf 421(6923)} 611--615.

\bibitem{kn:Ban71} Banks, E.R. (1971) Information and transmission in cellular automata, {\em PhD Dissertion}, Cambridge, MA, MIT.


\bibitem{kn:Benn73} Bennett, C.H. (1973) Logical reversibility of computation, {\em IBM Journal of Research and Development} {\bf 17(6)} 525--532.

\bibitem{ciappina2017attosecond} Ciappina, M.F., Hern\'andez, J.P., Landsman, A., Okell, W., Zherebtsov, S., F\"org, B., Sch\"otz, J., Seiffert, L., Fennel, T., Shaaran, T. \& Zimmermann, T. (2017) Attosecond physics at the nanoscale, {\em Reports on Progress in Physics} {\bf 80(5)} 054401.

\bibitem{kn:Codd68} Codd, E.F. (1968) {\em Cellular Automata}, Academic Press, Inc. New York and London.

\bibitem{kn:Cook04} Cook, M. (2004) Universality in Elementary Cellular Automata, {\em Complex Systems} {\bf 15(1)} 1--40.

\bibitem{goulielmakis2008single} Goulielmakis, E., Schultze, M., Hofstetter, M., Yakovlev, V.S., Gagnon, J., Uiberacker, M., Aquila, A.L., Gullikson, E.M., Attwood, D.T., Kienberger, R. and Krausz, F. (2008) Single-cycle nonlinear optics, {\em Science} {\bf 320(5883)} 1614--1617.

\bibitem{kn:CS11} Moore, C. \& Mertens, S. (2011) {\em The Nature of Computation}, Oxford University Press.


\bibitem{kn:FT01} Fredkin, E. \& Toffoli, T. (2002) Design Principles for Achieving High-Performance Submicron Digital Technologies. In: {\em Collision-Based Computing}, A. Adamatzky (ed), Springer, chapter 2, pages 27--46.

\bibitem{kn:FT82} Fredkin, E. \& Toffoli, T. (1982) Conservative logic, {\em Int. J. Theoret. Phys.} {\bf 21} 219--253.

\bibitem{kn:Hey98} Hey, A.J.G. (1998) {\em Feynman and computation: exploring the limits of computers}, Perseus Books.

\bibitem{kn:Hutt10} Hutton, T.J. (2010) Codd's self-replicating computer, {\em Artificial Life} {\bf 16(2)} 99--117.

\bibitem{kn:JSS01} Jakubowski, M.H., Steiglitz, K. \& Squier, R. (2001) Computing with Solitons: A Review and Prospectus, {\em Multiple-Valued Logic} {\bf 6(5-6)} 439--462. (also republished in \cite{kn:Ada02})

\bibitem{kn:LN90} Lindgren, K. \& Nordahl M. (1990) Universal Computation in Simple One-Dimensional Cellular Automata, {\em Complex Systems} {\bf 4} 229--318.

\bibitem{kn:MAA10} Mart{\'i}nez, G.J., Adamatzky, A., Sanz, R.A. \& Mora, J.C.S.T. (2010) Complex dynamic emerging in Rule 30 with majority memory, {\em Complex Systems} {\bf 18(3)} 345--365.

\bibitem{kn:MAA12} Mart{\'i}nez, G.J., Adamatzky, A. \& Sanz, R.A. (2012) Complex dynamics of elementary cellular automata emerging in chaotic rule, {\em Int. J. of Bifurcation and Chaos} {\bf 22(2)} 1250023-13.

\bibitem{kn:MAA13} Mart{\'i}nez, G.J., Adamatzky, A. \& Sanz, R.A. (2013) Designing Complex Dynamics with Memory, {\em Int. J. Bifurcation and Chaos} {\bf 23(10)} 1330035-131.

\bibitem{kn:MAC12} Mart{\'i}nez, G.J., Adamatzky, A., Chen, F. \& Chua, L. (2012) On Soliton Collisions between Localizations in Complex Elementary Cellular Automata: Rules 54 and 110 and Beyond, {\em Complex Systems} {\bf 21(2)} 117--142.

\bibitem{kn:MAM06} Mart{\'i}nez, G.J., Adamatzky, A. \& McIntosh, H.V. (2006) Phenomenology of glider collisions in cellular automaton Rule 54 and associated logical gates, {\em Chaos, Solitons and Fractals} {\bf 28} 100--111.

\bibitem{kn:MAM16} Mart{\'i}nez, G.J., Adamatzky, A., \& McIntosh, H.V. (2016) A Computation in a Cellular Automaton Collider Rule 110, In: {\em Advances in Unconventional Computing Volume 1: Theory}, A. Adamatzky (Ed.), Springer, chapter 15, 391--428.

\bibitem{kn:Marg84} Margolus, N.H. (1984) Physics-like models of computation, {\em Physica D} {\bf 10 (1-2)} 81--95.

\bibitem{kn:Marg98} Margolus, N.H. (1998) Crystalline Computation, In: {\em Feynman and computation: exploring the limits of computers}, A.J.G. Hey (Ed.), Perseus Books, 267--305.

\bibitem{kn:Marg02} Margolus, N.H. (2002) Universal Cellular Automata Based on the Collisions of Soft Spheres, In: {\em Collision-Based Computing}, A. Adamatzky (Ed.), Springer, chapter 5, 107--134.

\bibitem{kn:MAS10} Mart{\'i}nez, G.J., Adamatzky, A., Mora, J.C.S.T. \& Alonso-Sanz, R. (2010) How to make dull cellular automata complex by adding memory: Rule 126 case study, {\em Complexity} {\bf 15(6)} 34--49.

\bibitem{kn:MAS11} Mart{\'i}nez, G.J., Adamatzky, A., Stephens, C.R. \& Hoeflich, A.F. (2011) Cellular automaton supercolliders, {\em International Journal of Modern Physics C} {\bf 22(4)} 419--439.

\bibitem{kn:Mc90} McIntosh, H.V. (1990) Wolfram's Class IV and a Good Life, {\em Physica D} {\bf 45} 105--121.

\bibitem{kn:Mc09} McIntosh, H.V. (2009) {\em One Dimensional Cellular Automata}, Luniver Press.

\bibitem{kn:MH89} Morita, K. \& Harao, M. (1989) Computation universality of one-dimensional reversible (injective) cellular automata, {\em Trans. IEICE Japan} {\bf E-72} 758--762. 

\bibitem{kn:Mills08} Mills, J.W. (2008) The Nature of the Extended Analog Computer, {\em Physica D} {\bf 237(9)} 1235--1256.

\bibitem{kn:Mins67} Minsky, M. (1967) {\em Computation: Finite and Infinite Machines}, Prentice Hall.

\bibitem{kn:Mit01} Mitchell, M. (2001) Life and evolution in computers,  {\em History and Philosophy of the Life Sciences} {\bf 23} 361--383.

\bibitem{kn:MMS11} Mart{\'i}nez, G.J., McIntosh, H.V., Mora, J.C.S.T. \& Vergara, S.V.C. (2011) Reproducing the cyclic tag system developed by Matthew Cook with Rule 110 using the phases f$_1$\_1, {\em Journal of Cellular Automata} {\bf 6(2-3)} 121--161.

\bibitem{moore1970three} Moore, P.B., Huxley, H.E. \& DeRosier, D.J. (1970) Three-dimensional reconstruction of F-actin, thin filaments and decorated thin filaments, {\em Journal of Molecular Biology} {\bf 50(2)} 279--288.

\bibitem{kn:Mor90} Morita, K. (1990) A simple construction method of a reversible finite automaton out of Fredkin gates, and its related problem, {\em Trans. IEICE Japan} {\bf E-73} 978--984.

\bibitem{kn:Mor00} Morita, K. (2000) A new universal logic element for reversible computing, {\em IEICE technical report. Theoretical foundations of Computing} {\bf 99(724)} 119--126.

\bibitem{kn:Mor07} Morita, K. (2007) Simple universal one-dimensional reversible cellular automata, {\em Journal of Cellular Automata} {\bf 2} 159--165.

\bibitem{kn:Mor08} Morita, K. (2008) Reversible computing and cellular automata---A survey, {\em Theoretical Computer Science} {\bf 395} 101--131

\bibitem{kn:Mor16} Morita, K. (2016) Universality of 8-State Reversible and Conservative Triangular Partitioned Cellular Automata, {\em Lecture Notes in Computer Science} {\bf 9863} 45--54

\bibitem{kn:MSZ13a} Mart{\'i}nez, G.J., Mora, J.C.S.T. \& Zenil, H. (2013) Computation and Universality: Class IV versus Class III Cellular Automata, {\em Journal of Cellular Automata} {\bf 7(5-6)} 393--430.

\bibitem{kn:MTV86} Margolus, N., Toffoli, T. \& Vichniac, G. (1986) Cellular-Automata Supercomputers for Fluid Dynamics Modeling, {\em Physical Review Letters} {\bf 56(16)} 1694--1696.

\bibitem{nabekawa2017probing} Nabekawa, Y., Okino, T. \& Midorikawa, K. (2017) Probing attosecond dynamics of molecules by an intense a-few-pulse attosecond pulse train. In: {\em 31st International Congress on High-Speed Imaging and Photonics}, pp. 103280B. International Society for Optics and Photonics, Osaka, Japan.

\bibitem{kn:PST86} Park, J.K., Steiglitz, K. \& Thurston, W.P. (1986) Soliton-like behavior in automata, {\em Physica D} {\bf 19} 423--432.


\bibitem{kn:Ren16} Rendell, P. (2016) Turing Machine Universality of the Game of Life, {\em Springer}.

\bibitem{kn:Alo09} Sanz, R.A. (2009) {\em Cellular Automata with Memory}, Old City Publishing.

\bibitem{kn:Smi71} Smith III, A.R. (1971) Simple computation-universal cellular spaces, {\em J. of the Assoc. for Computing Machinery} {\bf 18} 339--353.

\bibitem{szent2004early} Szent-Gy\'{o}rgyi, A.G. (2004). The early history of the biochemistry of muscle contraction, {\em Journal of General Physiology} {\bf 123(6)} 631--641.

\bibitem{spudich1972regulation} Spudich, J.A., Huxley, H.E. \& Finch, J.T. (1972) Regulation of skeletal muscle contraction: II. Structural studies of the interaction of the tropomyosin-troponin complex with actin, {\em Journal of molecular biology} {\bf 72(3)} 619--632.

\bibitem{kn:SKW88} Steiglitz, K., Kamal, I. \& Watson, A. (1988) Embedding Computation in One-Dimensional Automata by Phase Coding Solitons, {\em IEEE Transactions on Computers} {\bf 37(2)} 138--145.

\bibitem{kn:Steig16} Steiglitz, K. (2016) Soliton-Guided Quantum Information Processing, In: {\em Advances in Unconventional Computing Volume 2: Prototypes, Models and Algorithms}, A. Adamatzky (Ed.), Springer, chapter 13, 297--307.

\bibitem{kn:Toff98} Toffoli, T. (1998) Non-Conventional Computers, In: {\em Encyclopedia of Electrical and Electronics Engineering}, J. Webster (Ed.), {\bf 14} 455--471, Wiley \& Sons.

\bibitem{kn:Toff02} Toffoli, T. (2002) Symbol Super Colliders, In: {\em Collision-Based Computing}, A. Adamatzky (Ed.), Springer, chapter 1, 1--22.

\bibitem{tuszynski2004ionic} Tuszy\'nski, J.A., Portet, S., Dixon, J.M., Luxford, C. \& Cantiello, H.F. (2004) Ionic wave propagation along actin filaments, {\em Biophysical Journal} {\bf 86(4)} 1890--1903.

\bibitem{kn:von66} von Neumann, J. (1966) {\em Theory of Self-reproducing Automata} (edited and completed by A. W. Burks), University of Illinois Press, Urbana and London.

\bibitem{kn:Wolf84} Wolfram, S. (1984) Universality and complexity in cellular automata, {\em Physica D} {\bf 10} 1--35.

\bibitem{kn:Wolf84a} Wolfram, S. (1984) Computation Theory of Cellular Automata, {\em Communications in Mathematical Physics} {\bf 96} 15--57.

\bibitem{kn:Wolf94} Wolfram, S. (1994) {\em Cellular Automata and Complexity}, Addison-Wesley Publishing Company.

\bibitem{kn:Wolf02} Wolfram, S. (2002) {\em A New Kind of Science}, Wolfram Media, Inc., Champaign, Illinois.

\end{thebibliography}
\end{document}